\documentclass{emulateapj}
\usepackage{apjfonts}
\newcommand{\nh}{$N_{\rm H}$}

\def\beq{\begin{equation}}
\def\enq{\end{equation}}
\def\begar{\begin{eqnarray}}
\def\endar{\end{eqnarray}}
\newcommand{\Msun}{\mbox{$M_{\odot}\;$}}
\def\lsim{\;\raise0.3ex\hbox{$<$\kern-0.75em\raise-1.1ex\hbox{$\sim$}}\;}
\def\gsim{\;\raise0.3ex\hbox{$>$\kern-0.75em\raise-1.1ex\hbox{$\sim$}}\;}

\def\beq{\begin{equation}}
\def\enq{\end{equation}}
\def\begar{\begin{eqnarray}}
\def\endar{\end{eqnarray}}
\def\mathnew{\mathsurround=0pt}
\def\simov#1#2{\lower .5pt\vbox{\baselineskip0pt \lineskip-.5pt
        \ialign{$\mathnew#1\hfil##\hfil$\crcr#2\crcr\sim\crcr}}}

\def\cmc{\rm ~cm^{-3}}
\def\cms{\rm ~cm^{-2}}
\def\kms{\rm ~km~s^{-1}}
\def\ergs{\rm ~ergs~s^{-1}}

\def\etal{{ et al. }}

\def\cmc{\rm ~cm^{-3}}
\def\cms{\rm ~cm^{-2}}
\def\diff{\rm ~cm^2~s^{-1}}
\def \kms {\rm ~km~s^{-1}}

\def\ergs{\rm ~ergs~s^{-1}}

\def\lfl{\rm ~ph~cm^{-2}~s^{-1} }
\def\enf{\rm ~ergs~cm^{-2}~s^{-1}}
\def\arcmin{\hbox{$^\prime$}}
\def\arcsec{\hbox{$^{\prime\prime}$}}
\def\efl{\hbox{ergs cm$^{-2}$ s$^{-1}$}}

\def \chan {{\sl Chandra}}
\def \xmm {{\sl XMM-Newton}}

\def \nh {$N_{\rm H}$}

\def \hcm {\hbox {\ifmmode $ atom cm$^{-2}\else atom cm$^{-2}$\fi}}
\def \arcmin {\hbox{$^\prime$} }
\def \arcsec {\hbox{$^{\prime\prime}$} }

\def \rchisq {$\chi_{\nu} ^{2}$}
\def\approxgt{\mathrel{\hbox{\rlap{\lower.55ex \hbox {$\sim$}}
        \kern-.3em \raise.4ex \hbox{$>$}}}}
\def\approxlt{\mathrel{\hbox{\rlap{\lower.55ex \hbox {$\sim$}}
        \kern-.3em \raise.4ex \hbox{$<$}}}}
\def\mathnew{\mathsurround=0pt}
\shorttitle{X-ray--IR sources in IC~443}
\shortauthors{A.M.Bykov et al.}

\begin{document}
\title{Isolated X-ray--infrared sources in
the region of interaction of the supernova remnant
IC~443 with a molecular cloud}

\author{A.M.\ Bykov\altaffilmark{1},
A.M.\ Krassilchtchikov\altaffilmark{1},
Yu.A.\ Uvarov\altaffilmark{1},
H.\ Bloemen\altaffilmark{2},
F.\ Bocchino\altaffilmark{3}, \\
G.M.\ Dubner\altaffilmark{4},
E.B.\ Giacani\altaffilmark{4},
G.G.\ Pavlov\altaffilmark{5}}
\altaffiltext{1}{A.F.\ Ioffe Institute for Physics and Technology,
St.\ Petersburg, Russia, 194021; byk@astro.ioffe.ru}
\altaffiltext{2}{SRON Netherlands Institute for Space Research,
Sorbonnelaan 2, 3584 CA Utrecht, The Netherlands}
\altaffiltext{3}{INAF -- Osservatorio Astronomico ``G.S.Vaiana",
Piazza del Parlamento 1, 90134 Palermo, Italy}
\altaffiltext{4}{Instituto de Astronom\'ia y F\'isica del Espacio
(IAFE), CC 67, Suc. 28, 1428 Buenos Aires, Argentina}
\altaffiltext{5}{Pennsylvania State University, 525 Davey
Laboratory, University Park, PA 16802}

\begin{abstract}
The nature of the extended hard X-ray source XMMU~J061804.3+222732
and its surroundings is investigated using {\sl XMM-Newton}, {\sl
Chandra}, and {\sl Spitzer} observations. This source is located
in an interaction region of the IC~443 supernova remnant with a
neighboring molecular cloud. The X-ray emission consists of a
number of bright clumps embedded in an extended structured
non-thermal X-ray nebula larger than 30\arcsec in size. Some
clumps show evidence for line emission at $\sim$ 1.9 keV and
$\sim$ 3.7 keV at the 99\% confidence level. Large-scale diffuse
radio emission of IC~443 passes over the source region, with an
enhancement near the source. An IR source of about 14\arcsec
$\times$ 7\arcsec size is prominent in the 24 $\mu$m, 70 $\mu$m,
and 2.2 $\mu$m bands, adjacent to a putative Si K-shell X-ray line
emission region. The observed IR/X-ray morphology and spectra
are consistent with those expected for J/C-type shocks of
different velocities driven by fragmented supernova ejecta
colliding with the dense medium of a molecular cloud. The IR
emission of the source detected by {\sl Spitzer} can be attributed
to both continuum emission from an HII region created by the
ejecta fragment and line emission excited by shocks. This source
region in IC~443 may be an example of a rather numerous population
of hard X-ray/IR sources created by supernova explosions in the
dense environment of star-forming regions. Alternative Galactic
and extragalactic interpretations of the observed source are also
discussed.
\end{abstract}

\keywords{ISM: individual (IC~443) --- supernova remnants --- X-rays: ISM}

\section{Introduction}

The energy release and the ejection of nucleosynthesis products
by supernovae (SNe) events are of great importance for our
understanding of the physics of the
interstellar medium (ISM). The mixing of the ejected metals with the
surrounding matter is of special interest when a SN occurs in a
molecular cloud, which may cause further star-forming activity.

Optical and UV studies of the structure of SN remnants (SNRs) have
revealed a complex metal composition of ejecta and the presence of
isolated high-velocity ejecta fragments interacting with surrounding
media. The most prominent manifestations of this phenomena are the
fast moving knots observed in some young ``oxygen-rich'' SNRs,
such as the Galactic SNRs Cas A  (e.g., Chevalier \& Kirshner 1979; Fesen\etal
2006), Puppis A (Winkler \& Kirshner 1985), G292.0+1.8 (e.g. Winkler \& Long 2006),
and also N132D in the LMC and 1E 0102.2--7219 in the SMC (e.g. Blair et al. 2000).

Ballistically moving ejecta fragments of SNRs can be considered as a
class of hard X-ray sources. The prototype was observed in the Vela
SNR (Aschenbach, Egger, \& Tr\"umper 1995; Miyata et al.\ 2001). A
massive individual fragment moving supersonically through a
molecular cloud can have a luminosity $L_x \gsim$10$^{31} \ergs$ in
the 1--10 keV band, and is observable with \xmm\ and \chan\ at a few
kpc distance (Bykov 2002, 2003). Its X-ray emission is expected to
consist of two components. The first one is thermal X-ray emission
from the hot shocked ambient gas behind the fragment bow shock, with
a spectrum of an optically thin thermal plasma of an ISM-cloud
abundance. The second emission component is nonthermal; the
interaction of fast electrons accelerated at the fragment bow-shock
with the fragment body produces a hard continuum as well as line
emission (X-ray and IR), including the K-shell lines of Si, S, Ar,
Ca, Fe, and other elements ejected by SN. Detection of the X-ray
line emission would help distinguish an ejecta fragment from the other
possible source of hard continuum emission associated with a SNR,
namely, a pulsar wind nebula (PWN).

A young SNR of an age of a few thousands years interacting with a
molecular cloud can produce hundreds of X-ray sources associated
with isolated ejecta fragments. They should be particularly numerous
in starforming regions like those in the Galactic center region,
where young core-collapsed supernovae in or near molecular clouds
are expected to be present in abundance.  The expected observational
appearance of isolated ejecta fragments in a molecular cloud differs
from what is seen in the Vela SNR. Ejecta fragments interacting with
a dense molecular cloud are slowed down and crushed, and they are
generally more bright. We will argue here that the X-ray emission
spectra of ejecta fragments in a molecular cloud may be dominated by
hard non-thermal components, because a powerful but very soft
thermal component could be heavily absorbed. On the other hand, the
spectra of fast supernova ejecta fragments propagating in a tenuous
plasma, as it is the case in the Vela SNR, would be long-lived, less
luminous and dominated by thermal emission.

 The present paper focuses on IC~443. This is a SNR of a medium
age, estimated by Chevalier (1999) to be  $\sim$ 30,000 years, for
which the number of X-ray sources from ejecta fragments should be
{\it much smaller\ } than in a young SNR (possibly, only a few). It is,
however, the best and most reliable laboratory to study this
phenomenon since there are only very few examples of clearly
established SNR-cloud interactions.

IC~443 (G189.1+3.0) is an evolved SNR of about 45$\arcmin$ size at a
distance of 1.5 kpc (e.g. Fesen \& Kirshner 1980). Radio
observations of IC~443 (e.g. Braun \& Strom 1986; Green 1986; Leahy
2004) show two half-shells.  This appearance is probably due to
interaction of the SNR with a molecular cloud that seems to separate
the two half-shells.  The molecular-cloud material has a
torus-like structure (Cornett, Chin \& Knapp 1977; Burton\etal
1988; Troja, Bocchino, \& Reale 2006), that can be interpreted
 as a sheet-like cloud first broken by the expanding pre-supernova
wind and then by the SNR blast wave.  Plenty of evidence for
shock-excited molecules in this region has been found
(e.g. DeNoyer 1979; Burton et al.\
1988; Dickman\etal 1992; Turner\etal 1992; van Dishoeck, Jansen,
\& Phillips 1993; Tauber\etal 1994; Richter, Graham, \& Wright
1995; Cesarsky\etal 1999; Snell\etal 2005). The complex structure
of the interaction region, with evidence for multiple dense clumps,
is seen in 2MASS images (e.g. Rho et al.\ 2001).
Three OH (1720 MHz) masers were found in IC~443
(Claussen \etal 1997; Hewitt \etal 2006, and references therein).

Soft X-ray maps of IC~443 based on {\sl ROSAT} data
(Asaoka \& Aschenbach 1994) and recent
radio observations (Leahy 2004) suggest that another SNR,
G189.6+3.3, is seen in the IC~443 field (see
also the \xmm\ study by Troja, Bocchino, \& Reale 2006).
This makes the multiwavelength observational picture even more complex to
interpret.

The field of IC~443 was observed in X-rays with {\sl HEAO 1\/}
(Petre et al.\ 1988), {\sl Ginga\/} (Wang et al.\ 1992), {\sl
ROSAT\/} (Asaoka \& Aschenbach 1994), {\sl ASCA\/} (Keohane et al.\
1997; Kawasaki et al.\ 2002), {\sl BeppoSAX\/} (Preite-Martinez et
al.\ 2000; Bocchino \& Bykov 2000), \chan\ (Olbert \etal 2001;
Bykov, Bocchino, \& Pavlov 2005; Gaensler \etal 2006; Weisskopf
\etal 2007), \xmm\ (Bocchino \& Bykov 2001, 2003; Troja, Bocchino, \& Reale 2006),
and {\sl RXTE} (Sturner, Keohane, \& Reimer 2004).

The X-ray emission of IC~443 below 4 keV is dominated by a number of
thermal components (e.g., Petre \etal 1988; Asaoka \& Aschenbach
1994; Kawasaki et al.\ 2002; Troja, Bocchino, \& Reale 2006). The
thermal-emission morphology is center-filled, with
soft emission filaments visible at energies below 0.5 keV.
A gradient of X-ray surface brightness at the SNR limb was found,
as well as strong variations of absorbing column density \nh,
which indicates the complex molecular-cloud environment of IC~443 in the
southern part of the remnant (e.g. Asaoka \& Aschenbach 1994).

\begin{figure*}
\includegraphics[width=0.99\textwidth]{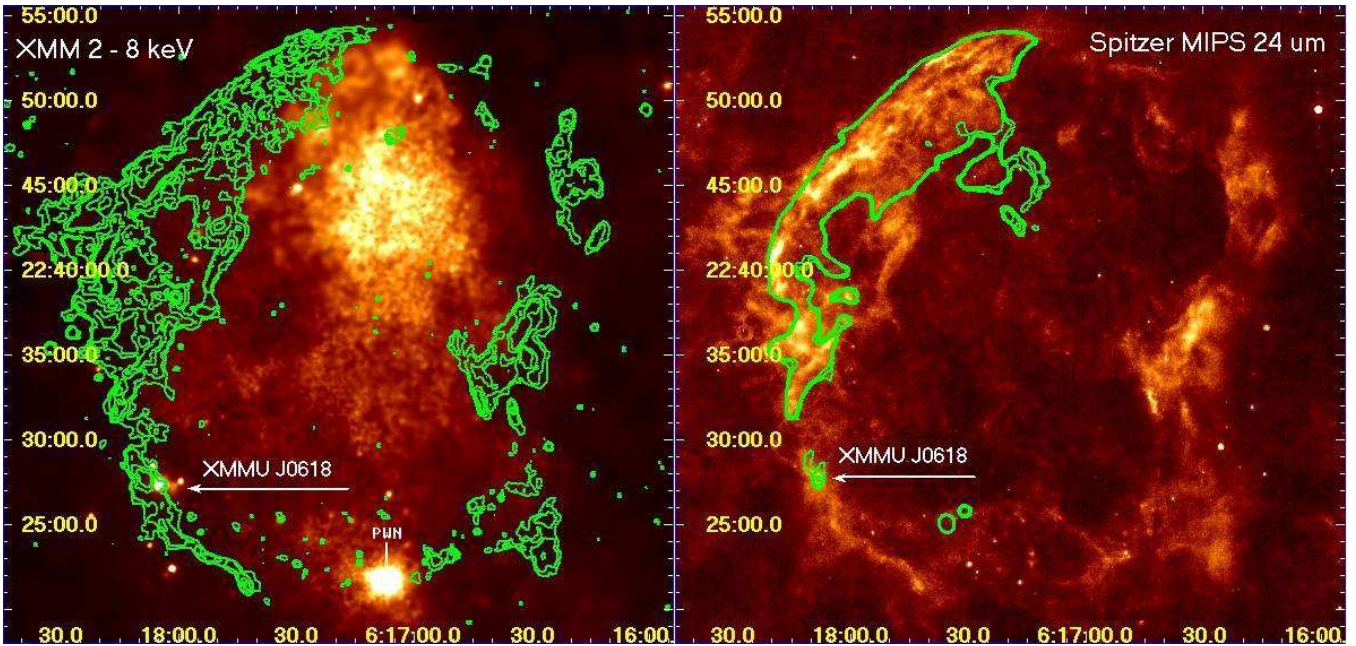}
\caption{ Wide-field views of SNR IC~443. {\em Left:} {\sl XMM-Newton} 2--8
keV image with {\sl Spitzer} MIPS 24 $\mu$m contours overlaid. {\em
Right:} {\sl Spitzer} MIPS image at 24 $\mu$m with {\sl VLA} 1.4 GHz
contours overlaid. The images are produced from the {\sl XMM-Newton}
observations 0114100101--0114100601 and 0301960101, and {\sl Spitzer} MIPS
observations r4616960, r4617216, and r4617472. The white
arrow points to the studied region.}
\label{xmm_spitzer}
\end{figure*}

 {\sl ASCA} observations have established that the hard X-ray
emission of IC~443 (above 4 keV) is dominated by localized sources
in the southern part of the remnant (Keohane et al.\ 1997). In \xmm\
observations Bocchino \& Bykov (2003; BB03 hereafter) found 12
sources with fluxes over 10$^{-14}~\enf$ in the 2--10 keV band. Six
of the detected sources are located in a relatively small, of
$15^\prime\times 15^\prime$ size, region projected onto the
molecular cloud in the South-Eastern part of IC~443. {\sl
BeppoSAX\/} MECS observations (4--10 keV) showed two sources,
1SAX~J0617.1+2221 and 1SAX~J0618.0+2227, with evidence from the {\sl
BeppoSAX\/} PDS for the presence of hard emission up to 100 keV for
the former (Bocchino \& Bykov 2000). Observations of this source by
\chan\ (Olbert\etal 2001; Gaensler\etal 2006; Weisskopf\etal 2007)
and \xmm\ (Bocchino \& Bykov 2001) established its plerionic nature.
Leahy (2004) argued that the pulsar that powers this plerion is
associated with G189.6+3.3 rather than IC~443. The nature of the
second hard source -- 1SAX~J0618.0+2227 -- remained unknown. This
source, the brightest in the region (excluding the plerion), was
resolved with \xmm\ into two sources -- the extended
XMMU~J061804.3+222732 (of $\sim$ 20\arcsec size) and the point-like
XMMU~J061806.4+222832. We will call them Src~1 and Src~2
respectively (note that the sources were listed as Src~11 and Src~12
in BB03). The position of XMMU~J061804.3+222732 in the remnant is
illustrated in Figures~\ref{xmm_spitzer} and \ref{chan_xmm}.
%\clearpage
%\clearpage
A dedicated \chan\ observation of Src~1 has revealed a complex
structure of a few bright clumps embedded in extended emission of
$>$ 20\arcsec size (Bykov, Bocchino, \& Pavlov 2005; BBP05
hereafter). The brightest clumps are the extended Src~1a and the
point-like Src~1b. The apparent position of the source in a SNR --
molecular cloud interaction region naturally leads to SNR-related
interpretations. The observed X-ray morphology of Src\,1 and the
spectra of its components are consistent with expectations for a SN
ejecta fragment interacting with a dense ambient medium.
Alternatively, Src\,1 could be interpreted as a PWN associated with
either IC~443 or G189.6+3.3 (BBP05). However, one cannot exclude the
extragalactic origin of the source, that is discussed in some detail
in Section~\ref{altern}.

IC~443 is a candidate counterpart of the EGRET  $\gamma$-ray source
3EG~J0617+2238, with a flux of about 5$\times$10$^{-7}$ cm$^{-2}$
s$^{-1}$ above 100 MeV (Esposito et al.\ 1996). The spectrum of
Src~1 extrapolated into the EGRET range is consistent with that of
3EG~J0617+2238 (BBP05). Also the position of Src~1 is consistent
(albeit marginally) with that of 3EG~J0617+2238. Such a $\gamma$-ray
luminosity can be expected for both the fragment and PWN
interpretations. The forthcoming {\sl GLAST\ } mission (e.g. Johnson
2006) will be able to provide an accurate position and spectrum of
3EG~J0617+2238, thus helping to solve the issue. Src~1 lies far away
from the 99\% error circle of the TeV-regime source recently
reported by {\sl MAGIC} (Albert et al. 2007) in the Western part of
IC~443 field. The apparent position of TeV {\sl MAGIC} source is
close to the 1720 MHz OH maser detected by Claussen et al. (1997).

We present here new results of a deep 80 ks
observation of the region with \xmm , imaging of the region with the
{\sl Spitzer} infrared observatory, and a new analysis of
VLA radio observations. In \S\,2 a combined analysis of the new \xmm\ observations
and all the previous high-resolution X-ray data from \chan\ and \xmm\ is
presented, including images, spectra, and time variations in the X-ray domain.
In \S\,3 archival radio (VLA), IR (2MASS and {\sl Spitzer} MIPS), and
optical (POSS-II) data are used to constrain the nature of Src~1.
A discussion of the obtained results and future prospects are presented in \S\,4.

\section{X-ray data analysis}

XMMU~J061804.3+222732 (Src~1) was observed with \xmm\ in 2000 and 2006, and
with \chan\ in 2000 and 2004 (Table~\ref{obs_log}).
CIAO v.3.0 with CALDB v.3.2.2 was used for \chan\ data processing,
SAS v.20060628\_1801-7.0.0 for \xmm\ data processing, and the
HEASOFT~6.1 suite, including XSPEC v.12.3, for spectral fitting.

In the course of \xmm\ data reduction, patterns 0--4 of EPIC-PN detector events and
patterns 0--12 of MOS events were used.
To filter out the periods of enhanced particle background,
we applied a standard method of soft flare
detection\footnote{http://heasarc.gsfc.nasa.gov/docs/xmm/abc/node7.html}.
The filtering did not significantly change the good time for the
data taken in 2000, but reduced it in the dataset of 2006.

In the \chan\ observation of the year 2000, Src~1 is offset 17$\farcm$6
from the ACIS aim point, on chip S4. To increase the
quality of the ACIS chip S4 data the standard procedure was used to create an event2 file from the
archival event1 file, including a run of the {\sl destreak} task. As
a result, the good time of the observation decreased from 11.5 to 9.6 ks.
An appropriate webscript\footnote{http://asc.harvard.edu/cal/ASPECT/fix\_offset/fix\_offset.cgi}
was used to correct the \chan\ event files for the systematic position offset.

%\clearpage
\begin{table*}[h!tb]
\label{powertable} \caption[]{
%        Summary of the observations. \label{obs}
        X-ray observations of XMMU~J061804.3+222732. \label{obs}
        }
\begin{center}
\begin{tabular}{llcccc} \tableline\tableline
 Obs ID & Observatory & Instrument & Date of observation & Exposure  & Good time\\
 & & & (YYYY-MM-DD)  &  (ks) & (ks) \\ \tableline
760 & Chandra & ACIS-S& 2000-04-10/11  & 11.5 & 9.6 \\
\hline
&     & PN   &            & 23.2 & 19.5   \\
0114100301 & XMM & MOS1 & 2000-09-27 & 25.6 & 25.1   \\
&     & MOS2 &            & 25.6 & 25.1   \\
\hline
4675 & Chandra & ACIS-S& 2004-04-12/13  & 58.4 & 56.2 \\
\hline
&     & PN   &            & 79.9 & 52.9   \\
0301960101 & XMM & MOS1 & 2006-03-30/31 & 81.6 & 67.6   \\
&     & MOS2 &            & 81.6 & 68.2   \\
\tableline
\end{tabular}
   \\ \rule{0mm}{5mm}
\end{center}
\label{obs_log}
\end{table*}
%\clearpage

\begin{figure*}
\includegraphics[width=0.98\textwidth]{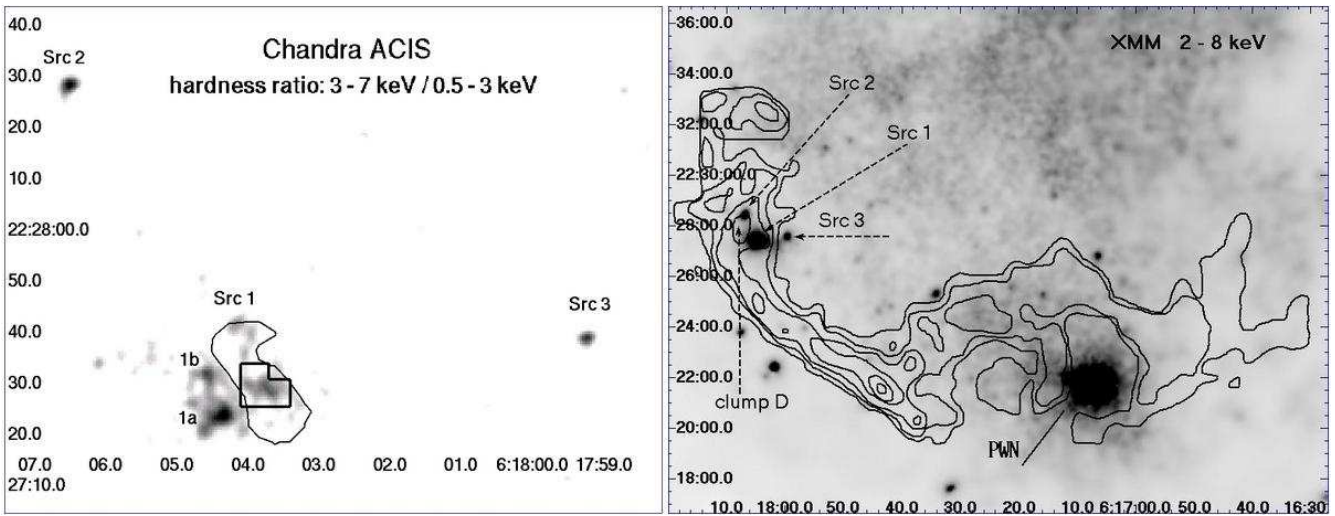}
\caption{X-ray images of the region under consideration. The data
are from the {\sl Chandra} observation 4675 and from \xmm\
observations 0114100101--0114100601 and 0301960101. {\em Left}: {\sl
Chandra} hardness ratio image (a 3.0--7.0 keV countrate map divided
by a 0.5--3.0 keV countrate map) together with a contour outlining
the extended near-IR source 2MASS~J06180378+2227314 (with the
L-shaped Si K-shell X-ray emission line region inside). Broad band
{\sl Chandra} maps of the region have been presented by BPP05.
 {\em Right}: \xmm\ image in the
 2--8 keV band with superimposed contours of 2.122 $\mu$m H$_2$ emission of Burton\etal (1987)
[adopted from van Dishoeck, Jansen, \& Phillips (1993)], indicating
the presence of a shocked molecular cloud. The studied X-ray sources
are indicated as well as the clump D of van Dishoeck, Jansen, \&
Phillips (1993). }
\label{chan_xmm}
\end{figure*}

 Figure~\ref{xmm_spitzer} shows a wide-field view of IC~443, both
in X-rays and in the medium IR 24 $\mu$m band. Two X-ray images of
the source region of interest here, obtained with \chan\ and \xmm\,,
are presented in Figure~\ref{chan_xmm}. The coordinates of Src~1a
and Src~1b obtained from the \chan\ data with the {\sl celldetect}
algorithm ($\alpha = 06^{\rm h}18^{\rm m}04\fs 32,
\delta +22^\circ 27' 24\farcs1$, and $\alpha = 06^{\rm h} 18^{\rm m} 04\fs
58, \delta = +22^\circ 27' 31\farcs8$, respectively) do not differ
from those reported by BBP05, as well as the position of Src~2
($\alpha = 06^{\rm h}18^{\rm m}06\fs 53, \delta = +22^\circ 28'
28\farcs2$). A nearby source, which we refer to as Src~3, is located
at $\alpha = 06^{\rm h}17^{\rm m}59\fs 25, \delta = +22^\circ 27'
38\farcs9$. In addition, the \xmm\ data show evidence for an
extended `bridge' of diffuse emission between Src~1 and Src~3 (see
Figures~\ref{xmm_spitzer}, \ref{chan_xmm}, and \ref{vla2}).  On the
timescale of about 6 years, covered by the \xmm\ observations used
here (Table~\ref{obs_log}), it is not possible to detect a proper
motion of the sources at a 1.5 kpc distance if they move with a
transverse velocity $\lsim$~3000~km s$^{-1}$.

%\clearpage

\begin{figure*}[!ht]
%\begin{tabular}{ccc}
\begin{tabular}{c}
\includegraphics[width=0.35\textwidth,bb=60 45 570 699,angle=270,clip]{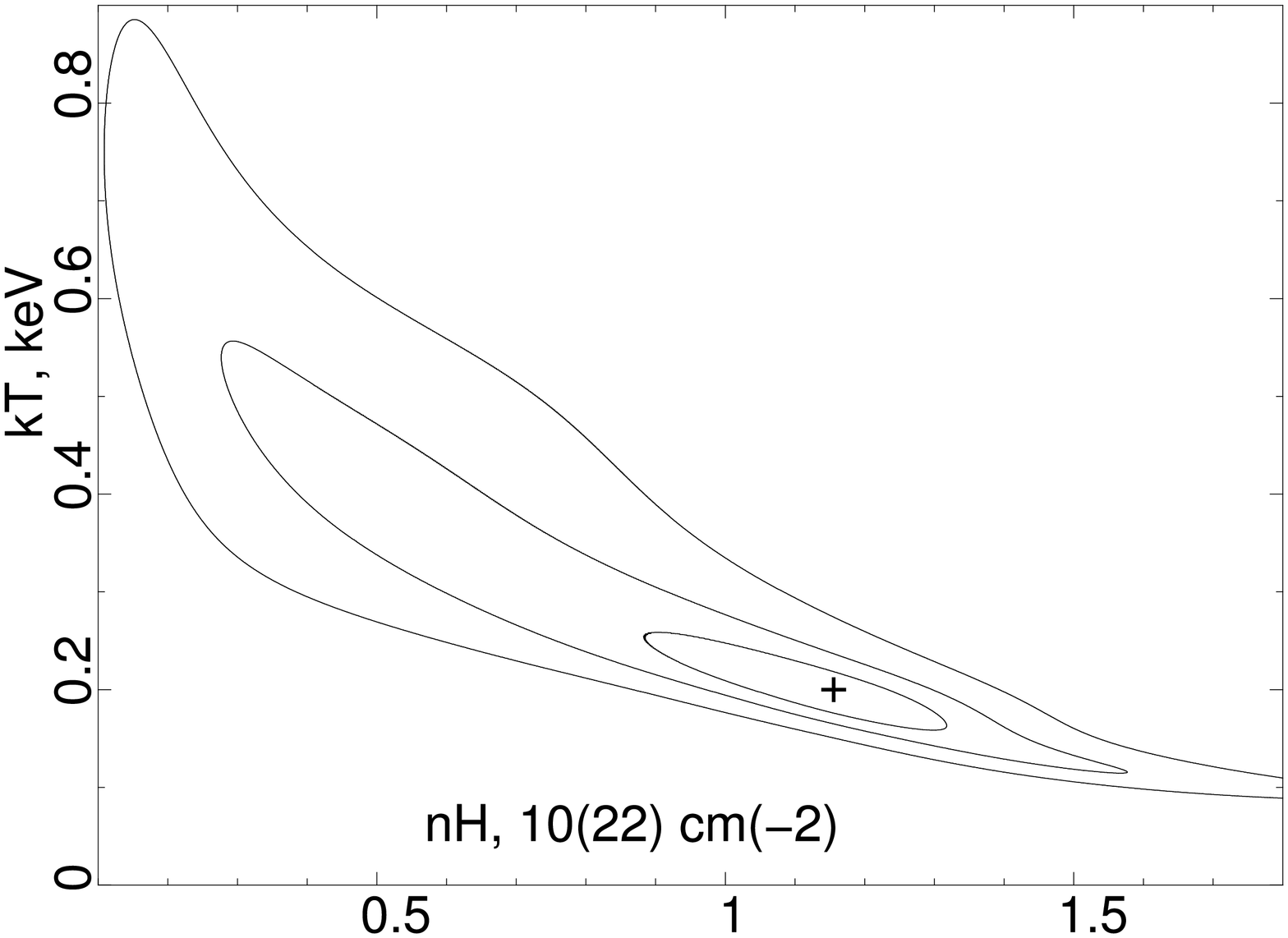} \\ % &
\includegraphics[width=0.35\textwidth,bb=60 45 570 699,angle=270,clip]{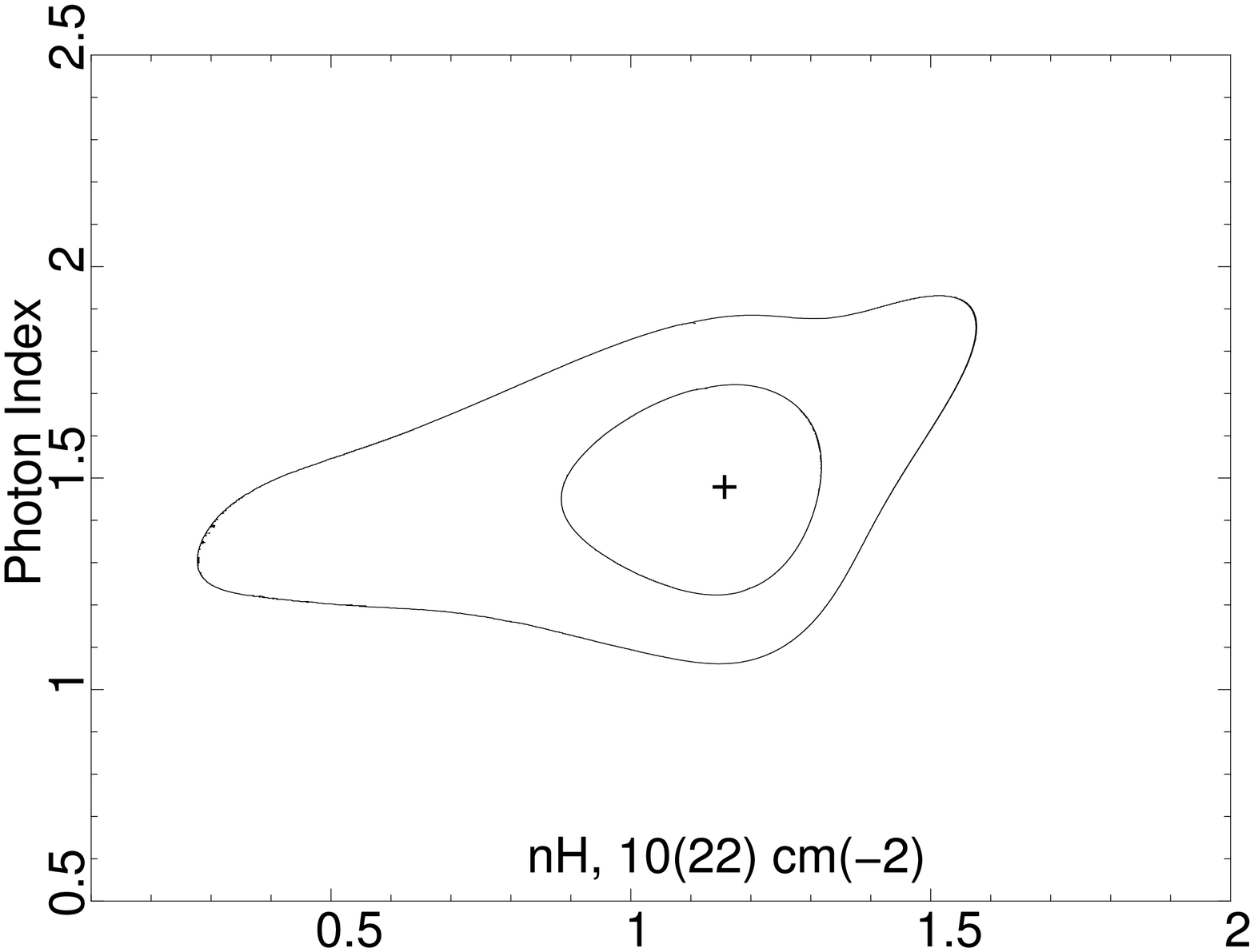} \\ % &
\includegraphics[width=0.35\textwidth,bb=60 45 570 700,angle=270,clip]{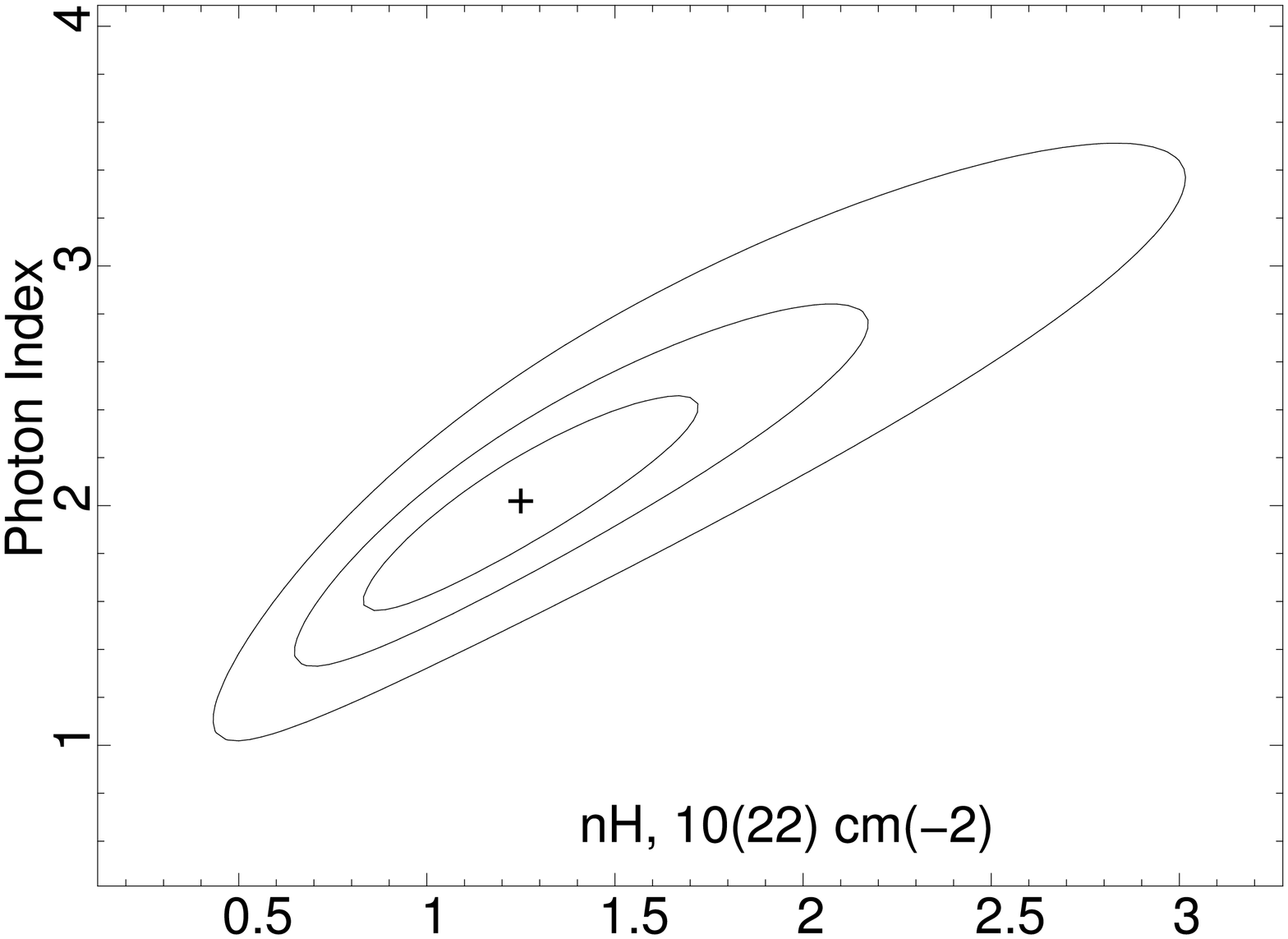} \\
\end{tabular}
\caption{ Confidence contours (68\%, 90\% and 99\% confidence
levels) for the spectral fit parameters of Src~1 spectra obtained
with \chan. {\em Upper and central:}  Temperature and photon index
vs.\ hydrogen column density for a two-component thermal plasma +
power law model. {\em Lower:}  Photon index vs.\ hydrogen column
density for an absorbed power law model. The outer contour on the central panel
runs outside the shown parameter space. } \label{fig:CL}
\end{figure*}

\begin{figure*}[!ht]
%\centering
\begin{tabular}{cc}
\includegraphics[width=0.35\textwidth,bb=46 30 583 777,angle=270,clip]{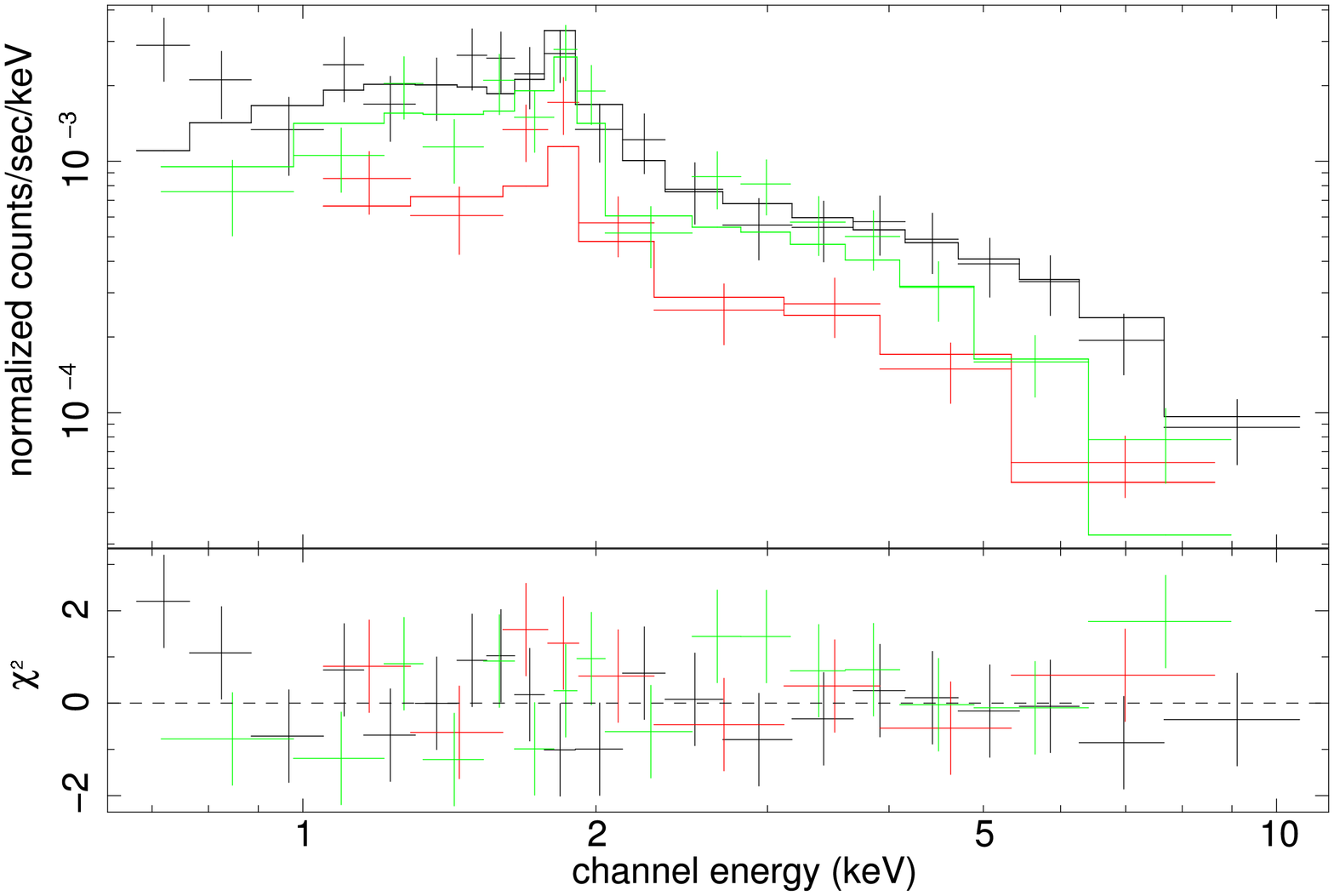} &
\includegraphics[width=0.35\textwidth,bb=34 44 579 724,angle=270,clip]{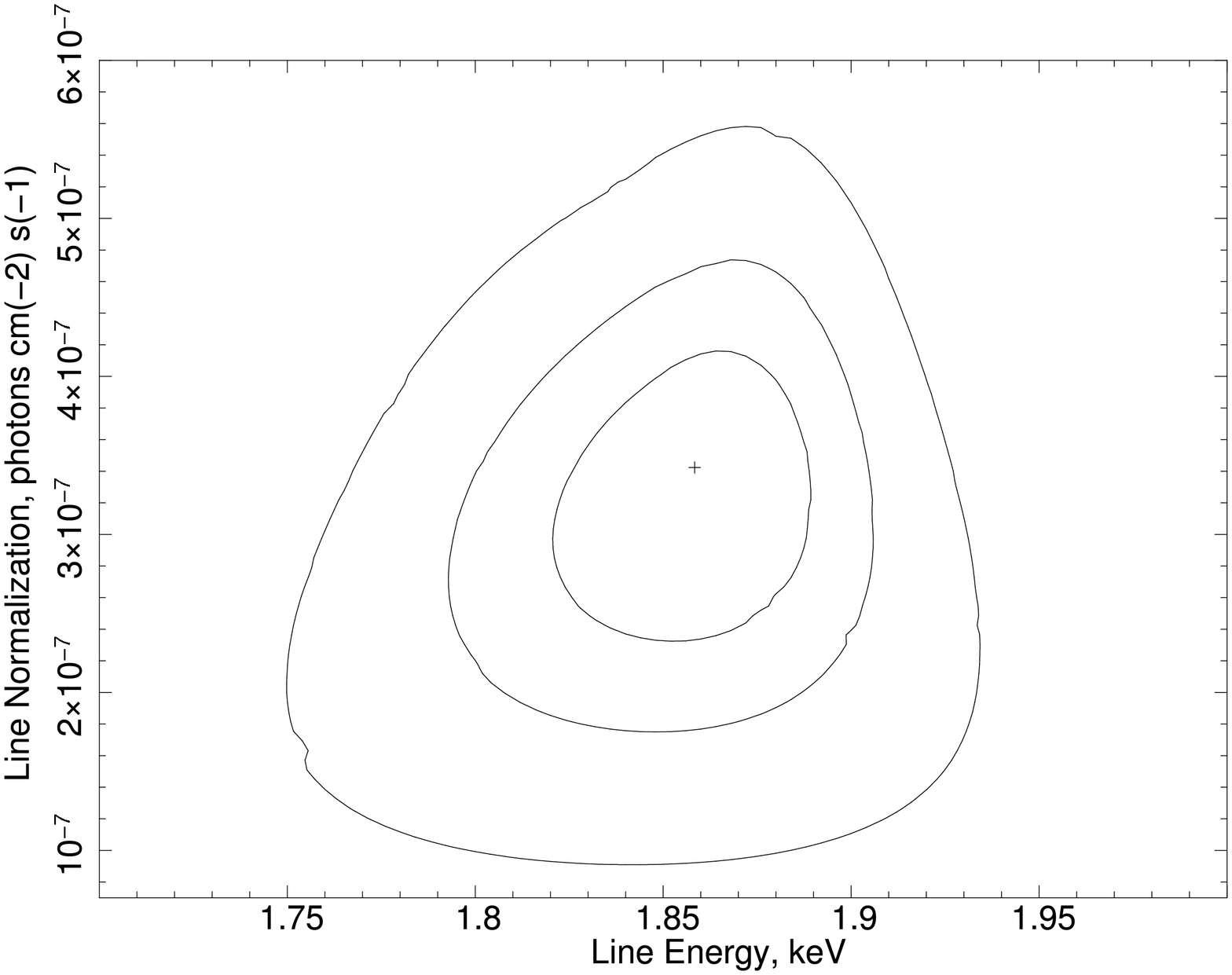} \\
\includegraphics[width=0.35\textwidth,bb=43 47 583 798,angle=270,clip]{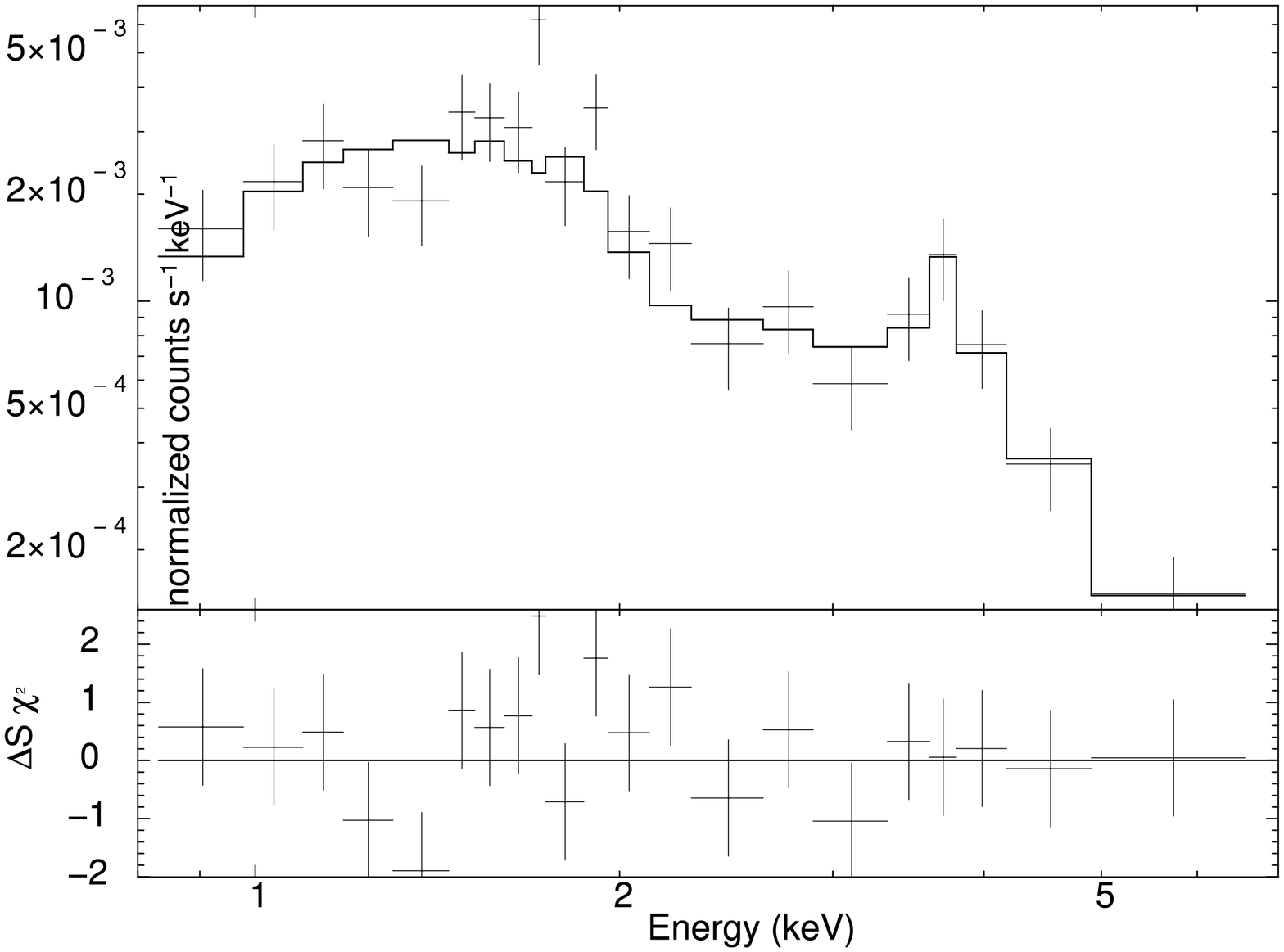} &
\includegraphics[width=0.35\textwidth,bb=62 12 587 711,angle=270,clip]{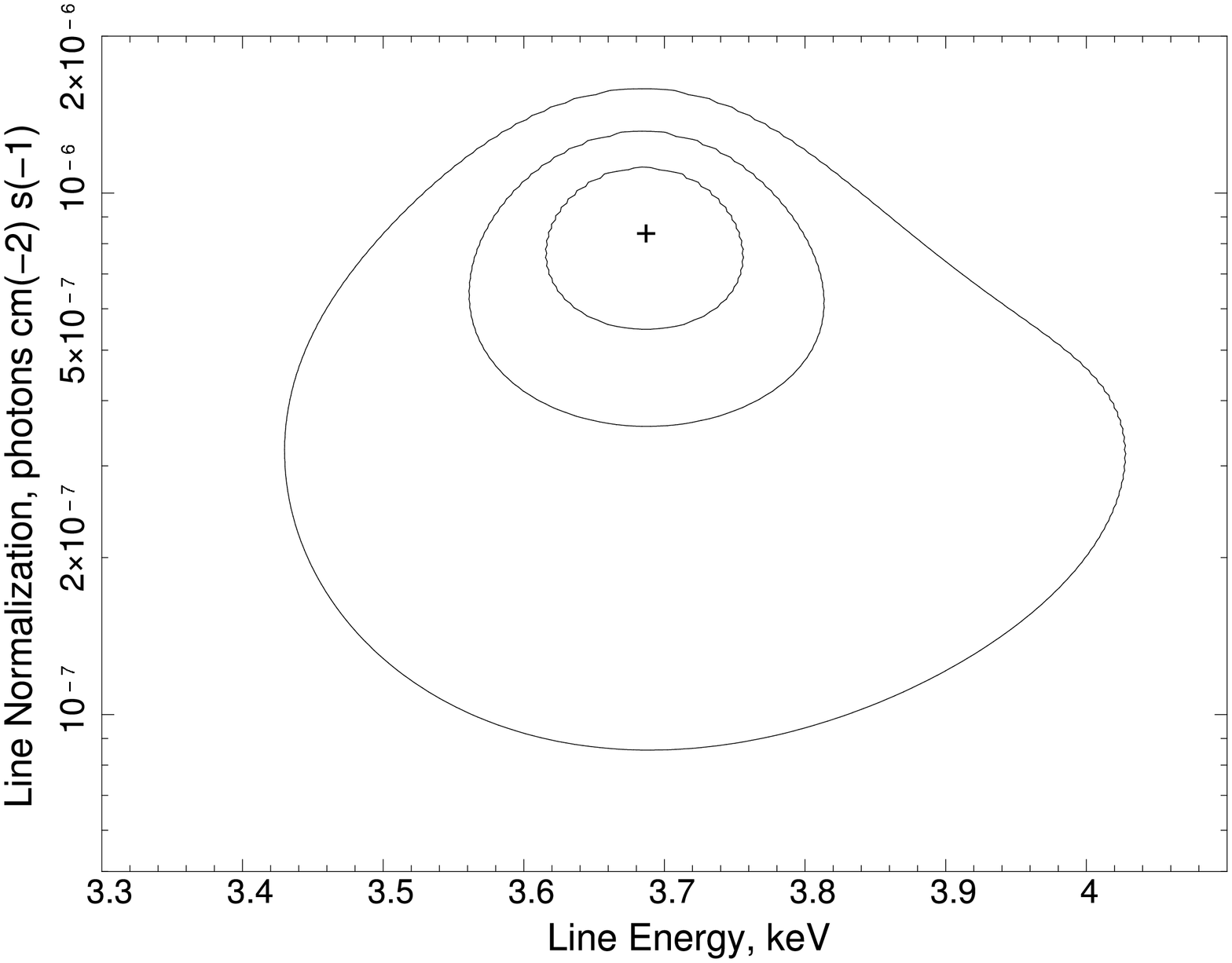} \\
\end{tabular}
\caption{
{\em Upper Left}: X-ray spectrum of the L-shaped region
(shown in Fig.\ref{chan_xmm}) derived from a combination of the MOS1, PN (obs. 2006), and ACIS (obs. 2004) data.
The model (histograms) corresponds to an absorbed [$N_H$=(0.3$\pm$0.2)$\times$10$^{22}$cm$^{-2}$]
power-law with $\Gamma$=1.1$^{+0.3}_{-0.2}$ and a possible Si line at 1.9$\pm$0.1~keV
of a fixed 0.1~keV width.
{\em Upper Right}: 68\%, 90\%, and 99\% confidence contours
for the parameters of the 1.8 keV Si K-shell line for the L-shaped region,
shown in Fig.\ref{chan_xmm}, obtained from combined data of the MOS1, PN (obs.
2006), and ACIS (obs. 2004) detectors.
{\em Lower Left}: X-ray spectrum of Src 3 extracted from
\chan\ observations. The spectrum is modeled as an absorbed
[$N_H$=0.6$^{+0.4}_{-0.2}\times$10$^{22}$cm$^{-2}$] power-law with
$\Gamma$=1.8$^{+0.4}_{-0.3}$ and a possible Ar line at
3.7$\pm$0.1~keV of a 0.2$^{+0.6}_{-0.2}$~keV width.
{\em Lower Right}: 68\%, 90\%, and 99\% confidence contours
for the parameters of the 3.7 keV Ar K-shell line for Src 3.
All the quoted errors are at the 90\% confidence level.
}
\label{fig:Ar}
\end{figure*}
%\clearpage

%*******************

\subsection{Source spectra}

The line of sight to the source region intersects the supernova
shell(s), the interior of the remnant, and the molecular
cloud, each having different physical parameters.
Therefore, the X-ray emission detected along the
line of sight has multiple thermal and nonthermal
components, in addition to the Galactic background.
A SNR shell can be characterized by a power-law emission
spectrum, the hot low-density interiors --- by thermal
plasma emission, and a dense molecular cloud may cause an appreciable absorption.

We have performed a series of spectral studies of Srcs~1, 2, and 3,
the results being presented in Table~\ref{pl_spec} (power-law
models), Table~\ref{tpl_spec} (thermal models), and in the text of
this subsection. {\it Mekal\ } thermal plasma models were used (e.g.
Kaastra 1992) and the {\it wabs} model for absorption was applied
(Morrison and McCammon 1983). Because of strong gradients in the
soft X-ray surface brightness, one should carefully choose regions
for spectral analysis. For the \chan\ analysis, circular source
regions of 10\arcsec radius surrounded by 10\arcsec--wide background
annuli were used. For \xmm, whose PSF is broader, source regions of
20\arcsec radius and annuli with 10\arcsec width were used. That
results in an underestimation of a source flux caused by subtraction
of the source photons regarded as background photons due to the wide
\xmm\ PSF wings. The effect is estimated to be about 20\% of the
source flux. That may account for the differences in the \chan\ and
\xmm\ normalizations of the fitted source flux reported in
Table~\ref{pl_spec}. However, the fitted power-law indices and
absorption values are not affected.

When comparing the \chan\ and \xmm\ spectra, such enlarged source
extraction regions were also used for the \chan\ data. For the
spectral fitting, the count rate spectra were grouped with a minimum
of 15 counts per bin. Table~\ref{pl_spec} summarizes the results of
spectral fitting for Src\,1, Src\,2, and Src\,3 with an absorbed
power-law (using the {\it wabs} model for absorption; Morrison and
McCammon 1983). Thermal plasma models  for Src\,1 and Src\,3 yield
very high and poorly constrained values of the temperature. However,
the spectrum of Src\,2 can be as well described with a {\it mekal}
model
%(e.g. Kaastra 1992)
as an absorbed emission of thermal plasma
(see Table~\ref{tpl_spec}).

The high angular resolution of \chan\ ACIS is useful for studying
the spectra of the substructure of Src~1 in more detail. A $2\arcsec
\times 4\arcsec$ elliptical source region around Src~1a, and a
2\arcsec radius source region around Src 1b were selected. For both
regions, the background counts were extracted from an annulus with
inner radius of 10\arcsec and outer radius of 20\arcsec
centered between the two sources. The spectrum of Src~1a can be
described as an absorbed ($N_H$ = (1.1$\pm$0.8)$\times$10$^{22}$
cm$^{-2}$)  power-law with photon index $\Gamma$ = 1.5$^{+0.5}_{-0.4}$
and a thermal plasma component with $T = 0.2^{+0.6}_{-0.1}$ keV.
The reduced $\chi^2$ of this fit is 0.71 at
14 d.o.f. The spectrum of Src~1b can be described as an absorbed
($N_H$~=~1.2$^{+1.0}_{-0.9}\times$10$^{22}$ cm$^{-2}$) power-law
with photon index $\Gamma$ = 2.0$^{+0.9}_{-0.7}$. The reduced
$\chi^2$ of this fit is 0.73 at 9 d.o.f. The quoted errors are at
the 90\% confidence level (for one interesting parameter). The
confidence contours for the spectral parameters of the whole Src~1
are shown in Figure~\ref{fig:CL}. These results are in a good
agreement with those obtained by BBP05.

The best-fit value of \nh\ for a power-law model of Src~1a (that
dominates the 10\arcsec radius Src~1 in Table~\ref{pl_spec}, line 4)
is well below that for the South-Eastern region of IC~443 obtained
with {\sl ROSAT} and \xmm\ (7$\times$10$^{21}$ cm$^{-2}$). If one
assumes that Src 1a, Src 1b and the extended hard emission are at
the same column density \nh $\sim$ 7$\times$10$^{21}$ cm$^{-2}$, one
has to add a soft thermal component (of T$\gsim$ 0.1 keV) for Src~1a
(See Fig.~3 in BBP05). Results of such fits, with the additional
optically thin thermal plasma emission ({\sl mekal}) component for
Src~1a, are presented in Table~1 of BBP05. Such a component is
localized in Src 1a and can be explained if the source is a
supersonic ejecta fragment (given a range of shock velocities up to
300 km/s).

Moreover, BBP05 showed that a satisfactory fit can be also obtained with
a blackbody component used instead of {\sl mekal}. Such a blackbody
plus power-law fit, with a temperature of about 0.1 keV, radius of about 3 km,
and power-law index of about 1.3 (at fixed \nh = 7$\times$10$^{21}$ cm$^{-2}$) can be
understood --- if Src~1a is a pulsar wind nebula --- as a combination of thermal emission
from the surface of a neutron star plus nonthermal emission from the neutron star magnetosphere
and/or surrounding pulsar wind nebula.

It should be noted that the high plasma temperatures obtained to fit
a thermal model to the \chan\ spectrum of the region of a 20\arcsec
radius around Src~1 are due to a contribution of hard spectra of
Src~1a, Src~1b and their neighbourhood. If one considers the
spectrum of this circle region with a removed inner circle of
11\arcsec radius surrounding  Src~1a and Src~1b (the resulting
annulus is named Src~1* in Table~\ref{tpl_spec}), one obtains a fit
with \nh = 1.1$^{+0.5}_{-0.4}\times$ 10$^{22}$ cm$^{-2}$,
T = 5.3$^{+11.4}_{-2.4}$ keV with $\chi^2$ = 0.93 at 78 d.o.f.

A line feature centered at $\approx 1.8$ keV is seen
(Figure~\ref{fig:Ar}, top row) at the 99\% confidence level in the
\xmm\ spectrum extracted from the extended nebula of Src~1 --- the
L-shaped region shown in the left panel of Fig.~\ref{chan_xmm}. The
combined MOS1 + PN + ACIS spectrum of the L-shaped region can be
modeled as an absorbed \nh=(0.3$\pm$0.2)$\times$10$^{22}$cm$^{-2}$]
power-law with $\Gamma$=1.1$^{+0.3}_{-0.2}$ and a possible Si line
at 1.9$\pm$0.1~keV of a fixed 0.1~keV width.

To investigate the \chan\ spectrum of the point-like Src~3, source
counts were extracted from a small circle of 2\arcsec radius. The
background counts were taken from an annulus with the inner radius
of 3\arcsec and the outer radius of 10\arcsec. The spectrum is shown
in Figure~\ref{fig:Ar} (bottom row). It contains a feature at 3.7
keV that is possibly due to an Ar emission line. There is, however,
an alternative possible interpretation of the line as a redshifted
Fe K line that assumes that Src~3 is extragalactic.

The \xmm\ spectrum of the faint bridge between Src~1 and Src~3 can
be modeled either as absorbed
(\nh=0.5$^{+0.3}_{-0.7}\times$10$^{22}$ cm$^{-2}$) thermal plasma (T
= 5.5$^{+3.7}_{-9.5}$ keV) emission with \rchisq = 0.79 at 74 d.o.f.
or as an absorbed (\nh=0.6$^{+0.4}_{-0.8}\times$10$^{22}$ cm$^{-2}$)
power-law of photon index $\Gamma$ = 2.0$^{+1.8}_{-2.1}$ with
\rchisq = 0.74 at 74 d.o.f.

The 20\arcsec--aperture \chan\ spectra of the studied sources Src~1,
Src~2 and Src~3 can be simultaneously modeled by an absorbed
(\nh=0.8$^{+0.2}_{-0.1}\times$10$^{22}$ cm$^{-2}$) power-law of the
photon index $\Gamma$ = 1.8$\pm$0.2 with \rchisq = 1.1 at 315 d.o.f.
or by a thermal plasma ({\sl mekal}) model with
\nh=(0.6$\pm$0.1$)\times$10$^{22}$ cm$^{-2}$, T=12$^{+7}_{-3}$ keV
with \rchisq = 1.1 at 315 d.o.f.

\begin{figure*}[t]
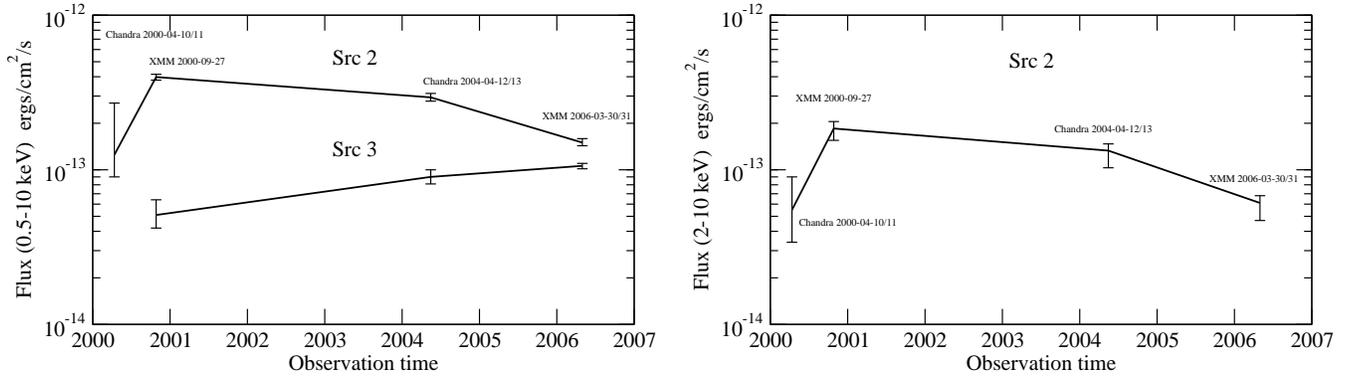

\begin{tabular}{cc}
\includegraphics[width=0.48\textwidth,bb=23 146 538 439 ,clip]{f5a.eps} &
\includegraphics[width=0.48\textwidth,bb=23 146 538 439 ,clip]{f5b.eps} \\
\end{tabular}
\caption{{\bf Left:} Fluxes of Src~2 and Src~3 as a function of time in the 0.5--10 keV
range. The errors are given at the 68\% confidence level.
{\bf Right:} Fluxes of Src~2 in the 2--10 keV range.}
\label{variab}
\end{figure*}

\begin{figure*}
\begin{tabular}{ll}
\includegraphics[width=0.55\textwidth,clip]{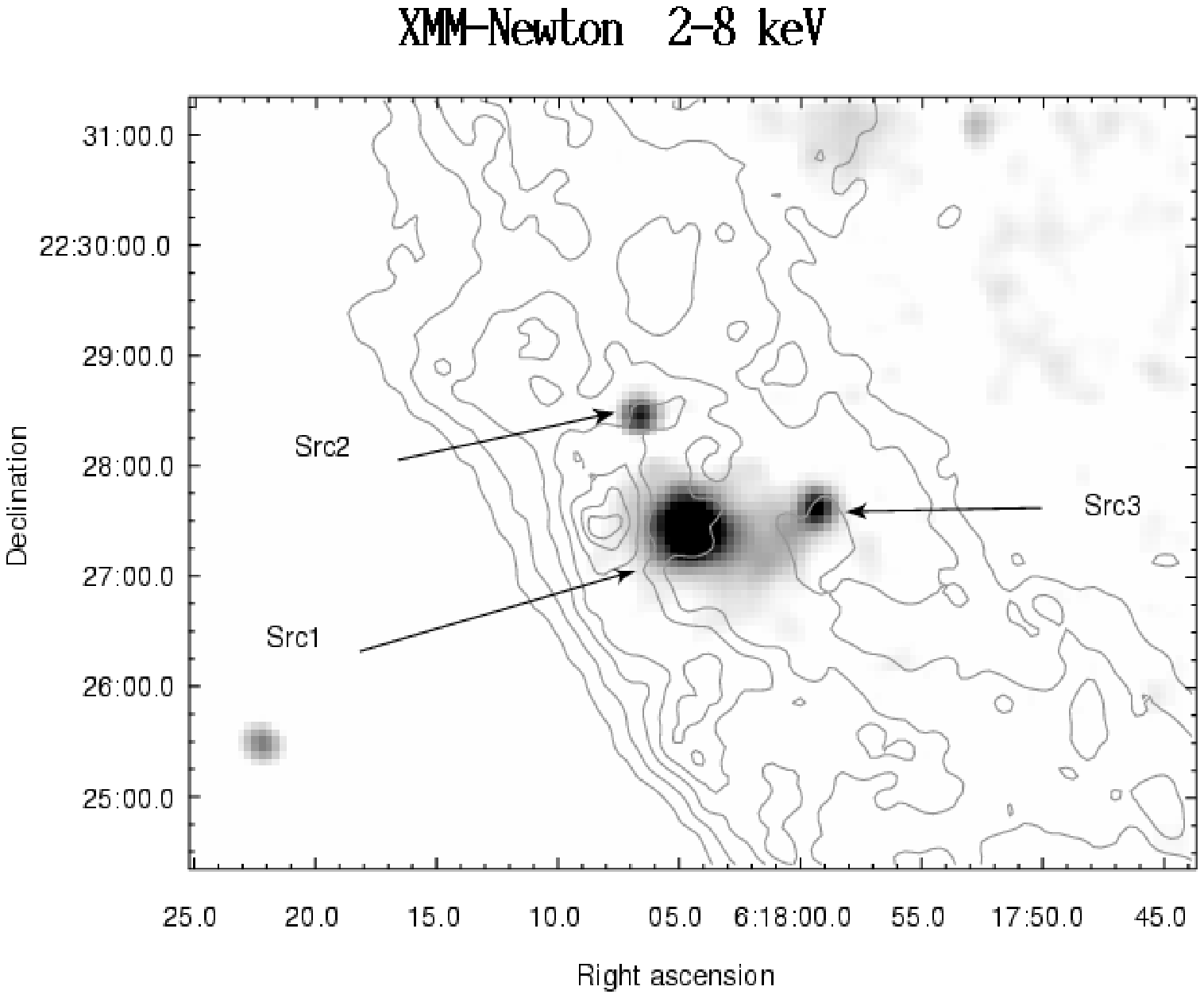} &
\includegraphics[width=0.45\textwidth,height=7.1cm,clip]{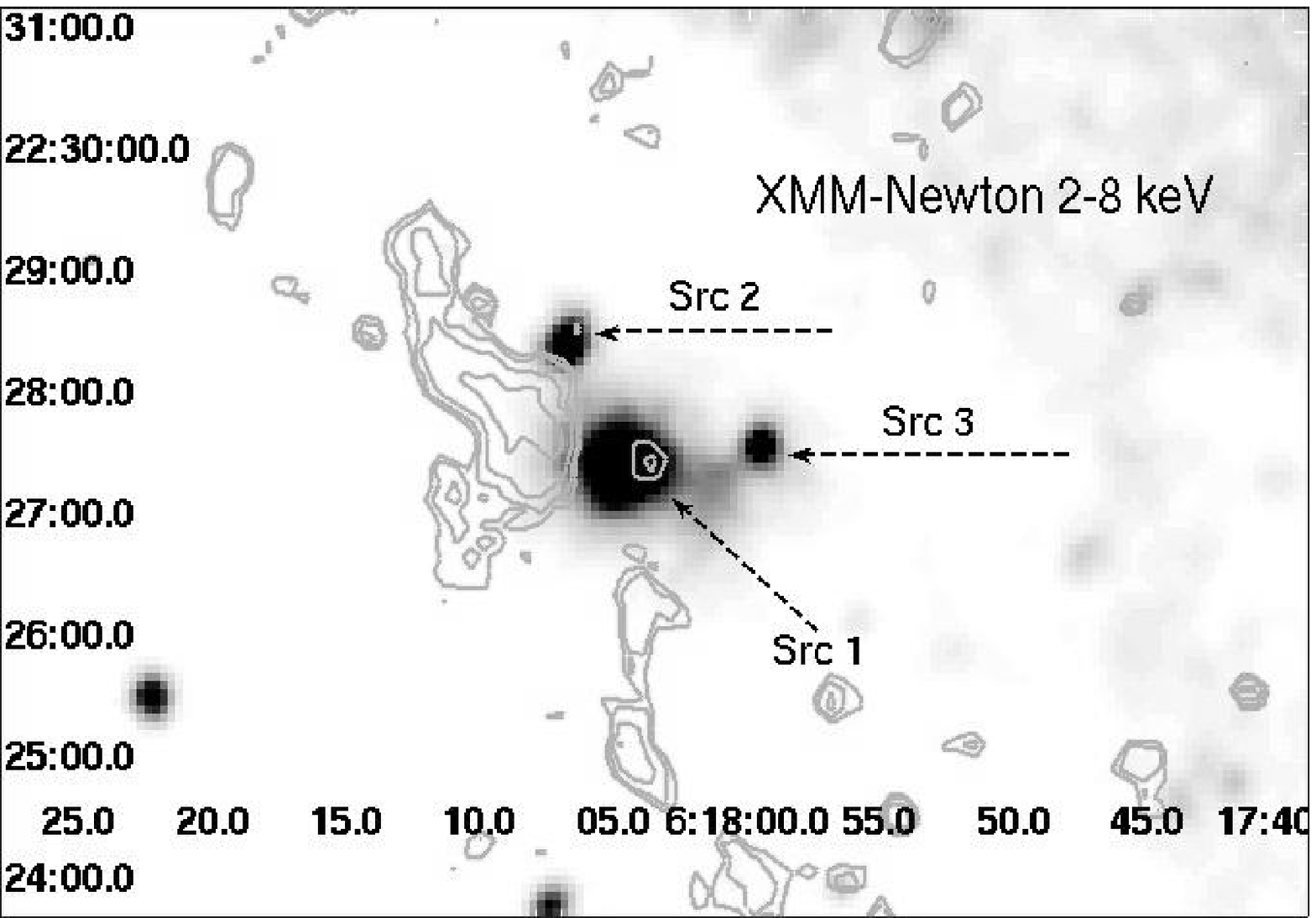} \\
\end{tabular}
\caption{{\bf Left:} Contours of radio surface brightness at
$\lambda$=20 cm superimposed on the XMM-Newton X-ray image in the
2-8 keV energy band. The angular resolution of the radio data
is 14$\farcs$1 $\times$ 13$\farcs$2. The radio contours are plotted at
1.2, 2.4, 3.6, 4.9, 6.1, 7.3, 8.0 and
8.6 mJy/beam. {\bf Right:} Contours of 24$\mu$m emission detected
by MIPS {\sl Spitzer} superimposed on the XMM-Newton X-ray image in
the 2-8 keV energy band. } \label{vla2}
\end{figure*}

Using all the data available, we found that Src~1 has not shown a significant
time variation, its flux being consistent with that originally
obtained by BB03. Some evidence (at the 90\% confidence level) is
found for time variation in the unabsorbed flux of Srcs~2 and 3,
as illustrated in Fig.~\ref{variab}.
The unabsorbed flux of Src 2 in the 0.5--10 keV band
decreased between 2000/09 and 2006/03 with 99\% confidence.
In the 2--10 keV band the flux increased between
2000/04 and 2000/09 and later decreased with 90\% confidence.
The unabsorbed flux of Src 3 increased between 2000/09 and 2006/03
with 90\% confidence.
More observations are needed to firmly conclude on the issue.

%\clearpage

%\clearpage

%\clearpage

\section{Radio, IR and optical data analysis}

\subsection{VLA data analysis}

A radio image of IC~443 was obtained from archival
VLA\footnote {The VLA of the NRAO is a facility of the NSF, operated
under cooperative agreement by Associated Universities, Inc.}
 data obtained in 1997 at 1465 MHz from observations  in the C and D arrays.
 The interferometric image has an angular resolution of 14$\farcs$1 $\times$ 13$\farcs$2.
To recover information at all spatial frequencies, the synthesis data were
combined with single dish data from the survey at 1408 MHz carried out
with the 100 m MPIfR telescope (Reich, Reich \& F\"urst 1990).
The final high fidelity image has an angular resolution
of 37$\farcs$8 $\times$ 34$\farcs$3,
PA = 46.5$^o$,
and an average rms noise $\sim 0.1$~mJy/beam.
The total integrated flux density over the whole IC~443, $S = (58 \pm 2)$ Jy,
is in a good agreement with the total integrated flux obtained from single dish observations
($S\sim 60$ Jy, Mufson et al.\ 1986), which assures
the accuracy of flux density estimates over selected portions of the SNR.

Figure~\ref{xmm_spitzer} (right panel) shows the contours of the $\lambda \sim$ 20~cm
emission of SNR IC~443 superimposed on a 24~$\mu$m {\sl
Spitzer} map.
The radio map for the source region of interest made from interferometric data only
is shown in Figure~\ref{vla2} as contours on a 2--8 keV \xmm\ X-ray image.
This figure shows that Src~1 lies at the periphery of IC~443, far
from  the main SNR radio shell  (that is situated in the North-East
of the remnant), but near a localized radio excess. At the angular
resolution and sensitivity of the present data, no radio continuum
source, either point-like or extended, could be associated with
any of the X-ray sources. The local radio flux density,
obtained by integrating the
radio emission over the region containing 97\% of the \xmm\ counts
of XMMU~J061804.3+222732, is ($60 \pm 3$) mJy.

\subsection{Spitzer MIPS imaging and photometry}

The field of IC~443 was the target of {\sl Spitzer} MIPS scan
observations \dataset[r4616960]{r4616960}, \dataset[r4617216]{r4616960},
and \dataset[r4617472]{r4616960} performed on
2005 November 9 (PI: G.\ Rieke). The Multiband Imaging Photometer for
{\sl Spitzer }(MIPS) aboard the {\sl Spitzer} Space Telescope
(Werner et al.\ 2004) is capable of imaging and photometry in broad
medium-IR spectral bands centered at 24$\,\mu$m, 70$\,\mu$m, and 160 $\,\mu$m,
and low-resolution spectroscopy between 55$\,\mu$m and 95$\,\mu$m
(Rieke et al.\ 2004). The 24$\,\mu$m band covers the range of 21.3$\,\mu$m -- 26.1$\,\mu$m,
the 70$\,\mu$m band covers the range of 61.5$\,\mu$m -- 80.5$\,\mu$m.
Srcs 1, 2, and 3 were outside the field of view of the {\sl Spitzer} IRAC
near-IR camera.

%\clearpage
\begin{figure*}[t]
\includegraphics[width=0.99\textwidth]{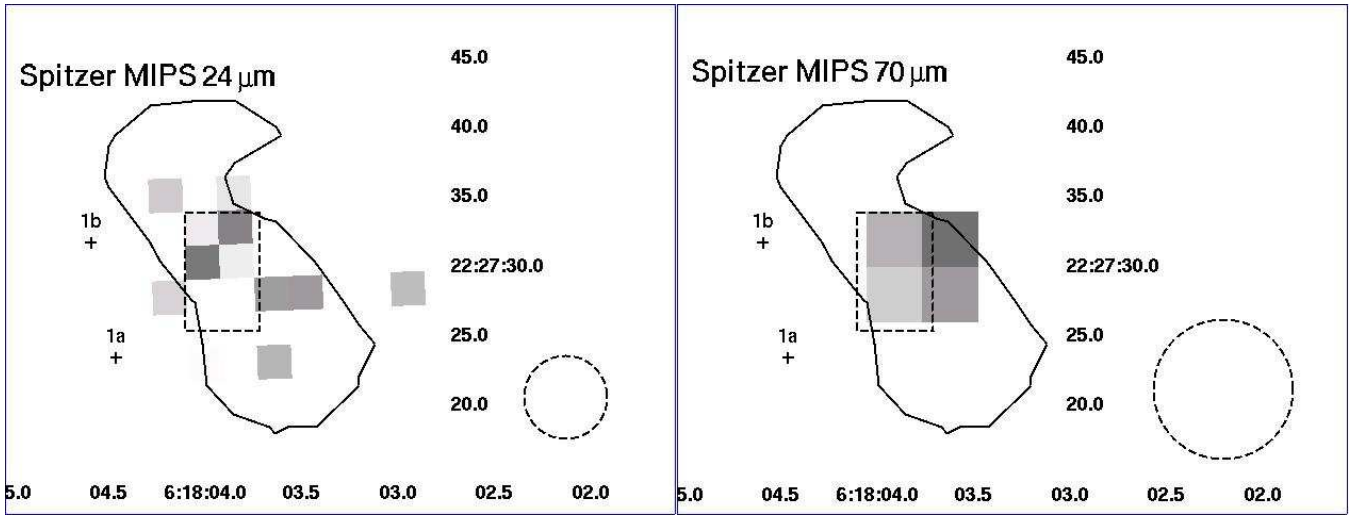}
\caption{ Src~1 environment as seen by {\sl Spitzer} MIPS. The
sources detected by {\sl Chandra} are shown as crosses (3--8 keV)
and a dashed rectangle (1.8--2.0 keV). The black contour denotes the
extended near-IR source 2MASS~J06180378+2227314. The dashed circles
denote the beamsize of MIPS. } \label{spit_map}
\end{figure*}
%\clearpage

We used the standard MOPEX~030106 software (Makovoz et al.\ 2006) to
construct mosaic images and extract point sources from the archival
BCD-level data (pre-processed by the S13 pipeline) according to the
recipes in the {\sl Spitzer}
cookbooks\footnote{http://ssc.spitzer.caltech.edu/documents/datademos/}
and the  MIPS data
handbook\footnote{http://ssc.spitzer.caltech.edu/mips/dh/mipsdatahandbook3.2.pdf}.
The first frames of each sequence were ignored. A total of 5940
individual frames were mosaiced for each of the MIPS bands. The net
exposure of the mosaic maps is equal to 65--92 frames (2.62 s each)
for the 24$\,\mu$m band and 12--15 frames (3.15 s each) for the
70$\,\mu$m band (different parts of the map were obtained with
different effective exposures). Outlier detection was performed to
exclude moving and solar system objects. A wide-field {\sl Spitzer}
MIPS image of the whole remnant is shown in
Figure~\ref{xmm_spitzer}.

Using the APEX software suite (Makovoz et al.\ 2006), we detected
two point-like sources,
at $\alpha = 06^{\rm h}18^{\rm m}04\fs 0, \delta = +22^\circ 27' 23$\arcsec
and
at $\alpha = 06^{\rm h}18^{\rm m}04\fs 0, \delta = +22^\circ 27' 33$\arcsec.
The sources were detected only in the 24$\,\mu$m band with fluxes
\hbox{1.77$\pm$0.07 mJy} (S/N = 8.2) and \hbox{1.89$\pm$0.07 mJy}
(S/N = 8.0), respectively. The latter source
coincides with the absorbed near-IR source 2MASS~J06180406+2227345 detected only in the
K$_s$ band ($J > 17.9$, $H > 17.0$, $K_s = 15.3\pm0.2$).
The dereddened K$_s$ (2.02 -- 2.30 $\mu$m) flux of the source
ranges from 0.64 to 1.21 mJy (considering the uncertainty of the
extinction value), that is (1.2--2.2)$\times$10$^{-13}$ \efl.

According to the extinction maps of Schlegel, Finkbeiner \& Davis
(1998), the total Galactic absorption towards Src~1 corresponds to
$A_{\rm V}\approx~6.0\pm 0.7$. This sets an upper limit because the
source is only $\sim$1.5 kpc away. Moreover, the method of Schlegel
et al.\ (1998) is known to overestimate the extinction for dense
regions by a factor of 1.3--1.5 (e.g., Arce \& Goodman 1999). Thus,
it is likely that $A_{\rm V}\sim 3$--4, consistent with the earlier
estimates by van Dishoeck, Jansen, \& Phillips (1993).

 There is another bright point-like source, 2MASS~J06180359+2227227 = HST~N8JT007783,
projected onto the area of Src~1. The source
is clearly seen in the optical and near-IR bands ($B = 16.3\pm 0.4$,
$V = 14.5\pm 0.3$, $J = 12.50\pm 0.03$, $H = 11.88\pm 0.03$, $K_s = 11.73\pm 0.02$).
The dereddened K$_s$ flux of the source ranges from
20.1 to 27.6 mJy, that is (3.6--5.0)$\times$10$^{-12}$ \efl. Most
likely, 2MASS~J06180359+2227227 is a foreground star.

An extended excess of IR emission is seen in the 24$\,\mu$m mosaic
map (left panel of Figure~\ref{spit_map}), coinciding with the
extended emission region detected with {\sl Chandra} ACIS in the
western part of Src~1 and with 2MASS~J06180378+2227314, an
extended (14\arcsec$\times$ 7\arcsec) source of near-IR emission
listed in the 2MASS XSC catalog with observed
isophotal\footnote{http://www.ipac.caltech.edu/2mass/releases/allsky/doc/sec4\_5e.html}
K$_s$~=~12.86$\pm$0.12. The excess is also seen in the 70$\,\mu$m
band (right panel of Figure~\ref{spit_map}), its apparent size
being comparable with the beam size. The aperture photometry
estimates of the excess (with aperture corrections applied) are
$11.4^{+1.9}_{-1.4}$~mJy (90\% err.) $\pm$ 1.1~mJy (3$\sigma$
error of the pipeline) for the 24$\,\mu$m band and
840$^{+170}_{-480}$~mJy (90\% err.) $\pm$ 170~mJy (3$\sigma$ error
of the pipeline) for the 70$\,\mu$m band. The apertures of a
6\arcsec and 8\arcsec radius were used for the 24$\,\mu$m and
70$\,\mu$m bands, respectively.
 These values correspond to
(2.8$^{+0.48}_{-0.35}\pm0.28$)$\times$10$^{-13}$ \efl\ for the
24$\mu$m band and (9.7$^{+1.96}_{-5.52}\pm 1.96$)$\times$10$^{-12}$
\efl\ for the 70$\mu$m band. Notice, that the latter value is actually an
upper limit.
Depending on the extinction value, the dereddened flux of
2MASS~J06180378+2227314 is 6.4--10.7 mJy, that is
(1.2--1.9)$\times$10$^{-12}$ \efl\ in the K$_s$ band.

The extended source is seen neither in the J and H bands of the
2MASS survey (Skrutskie et al.\ 2006) nor in the archival blue, red,
and infrared images of the POSS-II survey (Reid et al.\ 1991).
With the {\it daophot} package (Stetson 1987) integrated into the NOAO IRAF
software suite, the following upper limits were obtained for
2MASS~J06180378+2227314: 6.6$\times$10$^{-13}$ \efl\
in the 2MASS J band, 2.9$\times$10$^{-13}$ \efl\ in the 2MASS H band,
8.0$\times$10$^{-11}$ \efl\ in the POSS-II blue band  (3750--5500 \AA),
3.0$\times$10$^{-12}$ \efl\ in the POSS-II red band (5900--7100 \AA),
and 3.0$\times$10$^{-11}$ \efl\ in the POSS-II IR band (7350--8750 \AA).
The limits assume
the estimated extinction for the source ($A_{\rm V} = 6$).

\begin{figure*}
\centering
\includegraphics[width=0.8\textwidth]{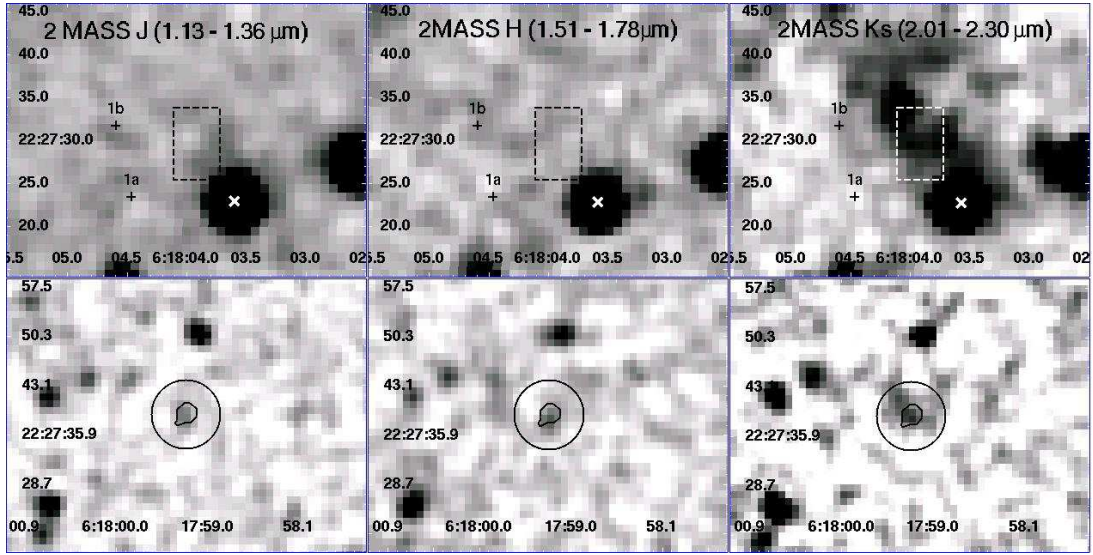}
\caption{{\em Upper panels:} Src~1 environment in the 2MASS near-IR
bands. The source regions and positions are marked in the same
manner as in Figure~\ref{spit_map}. The white X-mark denotes the
position of 2MASS~J06180359+2227227.  {\em Lower panels:}
Surroundings of Src~3 in the 2MASS near-IR bands. The 10\arcsec
radius circle denotes the region used for spectral analysis of the
X-ray data. The contour inside the circle denotes the position of
Src~3 as seen by \chan\ in the 0.3 -- 10 keV band at the
2.5$\times$10$^{-5}$ cps/pix level.} \label{opt_2mass}
\end{figure*}

The point-like X-ray source Src~3 is also seen as a weak IR source in all
the bands of the 2MASS survey, with
$J = 17.3\pm0.3$, $H = 16.8\pm0.6$, $K_s = 16.2\pm0.4$ (see lower panels of Figure~\ref{opt_2mass}).

%\clearpage

%\clearpage

\begin{deluxetable*}{ccccccc}
%\rotate 
%\tablewidth{0pt} 
\tablewidth{0.95\textwidth} 
\tablecaption{Parameters of absorbed
power-law models of Srcs 1, 2, and 3
\label{pl_spec}} 
\tablehead{
\multicolumn{2}{c}{ } & \\
% \colhead{Spectral parameters}\\
%\tableline 
%\colhead{Source(radius)/Background} &
%\colhead{Observatory, year} & \colhead{Source counts} &
%\colhead{$N_H$\tablenotemark{a}} & \colhead{$\Gamma$} &
%\colhead{$Norm$\tablenotemark{b}} & \colhead{$\chi^2_\nu$/dof} } 
}
\startdata 
 & & & & & & * \\
 & & & & & & * \\
 Source(radius)/Background  & Observatory, year & Source counts & 
 $N_H$\tablenotemark{a} & $\Gamma$ & $Norm$\tablenotemark{b} & $\chi^2_\nu$/dof \\
 & & & & & & * \\
\tableline 
Src 1 (20\arcsec)/bkg2 & XMM 2000 A&  1064/505/507 &
$0.5^{+0.3}_{-0.2}$ & $1.2^{+0.3}_{-0.2}$ &
$2.8^{+1.4}_{-0.9}10^{-5}$ &
1.13/124 \\
Src 1 (20\arcsec)/bkg2 & XMM 2006 A&  4574/1899/2150 &
$0.5^{+0.1}_{-0.1}$ & $1.3^{+0.1}_{-0.04}$ &
$3.2^{+0.6}_{-0.5}10^{-5}$ &
1.05/418 \\
Src 1 (20\arcsec)/bkg2 & Chandra 2004 & 2973 & $0.7^{+0.2}_{-0.1}$ &
$1.5^{+0.2}_{-0.2}$ & $6.7^{+2.2}_{-1.4}10^{-5}$ &
0.84/151 \\
Src 1 (10\arcsec)/bkg1 & Chandra 2004  & 1314 & $0.2^{+0.1}_{-0.1}$
& $1.0^{+0.2}_{-0.2}$ & $1.4^{+0.4}_{-0.3}10^{-5}$ &
0.97/73 \\
\tableline Src 2 (20\arcsec)/bkg2 & XMM 2000 A&  955/418/470 &
$0.8^{+0.2}_{-0.2}$ & $2.3^{+0.3}_{-0.1}$ &
$9.4^{+3.9}_{-2.5}10^{-5}$ &
0.94/107 \\
Src 2 (20\arcsec)/bkg2 & XMM 2006 B&  --/944/1003 &
$1.0^{+0.5}_{-0.3}$ & $2.7^{+0.5}_{-0.3}$ &
$5.2^{+4.6}_{-2.0}10^{-5}$ &
0.88/105 \\
Src 2 (20\arcsec)/bkg2 & Chandra 2004 & 1829 & $1.2^{+0.3}_{-0.2}$ &
$2.6^{+0.3}_{-0.3}$ & $12.2^{+7.2}_{-3.8}10^{-5}$ &
0.98/94 \\
Src 2 (10\arcsec)/bkg1 & Chandra 2004  & 1270 & $0.9^{+0.2}_{-0.2}$
& $2.4^{+0.2}_{-0.2}$ & $9.3^{+3.7}_{-2.3}10^{-5}$ &
1.05/66 \\
\tableline Src 3 (20\arcsec)/bkg2 & XMM 2000 B&  --/235/225 &
$0.5^{+3.0}_{-0.5}$ & $2.1^{+2.8}_{-1.5}$ &
$1.0^{+19}_{-1.0}10^{-5}$ &
1.15/26 \\
Src 3 (20\arcsec)/bkg2 & XMM 2006 A&  2730/940/1080 &
$0.6^{+0.6}_{-0.4}$ & $1.6^{+0.4}_{-0.3}$ &
$1.1^{+0.9}_{-0.4}10^{-5}$ &
1.10/234 \\
Src 3 (20\arcsec)/bkg2 & Chandra 2004 & 1251 & $0.3^{+0.5}_{-0.3}$ &
$1.5^{+0.6}_{-0.6}$ & $1.0^{+1.2}_{-1.0}10^{-5}$ &
1.10/66 \\
Src 3 (10\arcsec)/bkg1 & Chandra 2004  & 620 & $0.7^{+0.4}_{-0.2}$ &
$1.9^{+0.4}_{-0.3}$ & $1.7^{+1.3}_{-0.7}10^{-5}$ & 1.36/34
\enddata
\tablenotetext{a}{$N_H$ is in units of 10$^{22}$~cm$^{-2}$.}
\tablenotetext{b}{Normalization parameter is the spectral flux in
photons cm$^{-2}$ s$^{-1}$ keV$^{-1}$ at 1~keV.}
\tablecomments{`A' means PN, MOS1,
MOS2 data combined, while `B' means MOS1 and MOS2 data combined;
`bkg1' denotes a background annulus with inner radius of
10\arcsec and outer radius of 20\arcsec, `bkg2' denotes a background
annulus with inner radius of 20\arcsec and outer radius of
30\arcsec. The source counts in the third column are given as
PN/MOS1/MOS2 for XMM and ACIS for Chandra. All errors quoted in the
table are at the 90\% confidence level for one interesting
parameter.}
\end{deluxetable*}

%\clearpage

\begin{deluxetable*}{ccccccc}
%\rotate 
%\tablewidth{0pt} 
\tablewidth{0.95\textwidth} 
\tablecaption{Parameters of absorbed plasma
(mekal) models of Srcs 1, 2, and 3\label{tpl_spec}} \tablehead{
\multicolumn{2}{c}{ } & \\
% \colhead{Spectral parameters}\\
%\tableline \colhead{Source(radius)/Background} &
%\colhead{Observatory, year} & \colhead{Source counts} &
%\colhead{$N_H$\tablenotemark{a}} & \colhead{$T, keV$} &
%\colhead{$Norm$\tablenotemark{b}} & \colhead{$\chi^2_\nu$/dof} }
}
\startdata
 & & & & & & * \\
 & & & & & & * \\
Source(radius)/Background & Observatory, year & Source counts  &
$N_H$\tablenotemark{a} & $T, keV$ & $Norm$\tablenotemark{b} & $\chi^2_\nu$/dof \\
 & & & & & & * \\
 \tableline
Src 1 (20\arcsec)/bkg2 & XMM 2000 A&  1064/505/507 &
$0.5^{+0.2}_{-0.1}$  & $79.9^{+0}_{-59.6}$&
$1.9^{+0.2}_{-0.4}10^{-4}$ &
1.1/124 \\
Src 1 (20\arcsec)/bkg2 & XMM 2006 A&  4574/1899/2150 &
$0.4^{+0.1}_{-0.1}$ & $75.1^{+4.8}_{-37.6}$ &
$1.8^{+0.1}_{-0.2}10^{-4}$ &
1.1/418 \\
Src 1 (20\arcsec)/bkg2 & Chandra 2004 & 2973 &
$0.6^{+0.1}_{-0.1}$ & $24.6^{+55.3}_{-10.8}$ &
$2.7^{+0.5}_{-0.2}10^{-4}$ &
0.85/151 \\
Src 1 (10\arcsec)/bkg1 & Chandra 2004  & 1314 &
$0.3^{+0.1}_{-0.1}$ & 79.8 $^{+0.1}_{-50.4}$ &
$1.1^{+0.1}_{-0.2}10^{-4}$ &
0.93/73 \\
\tableline

Src 1* (20\arcsec)/bkg2 & Chandra 2004 & 1781 &
$1.1^{+0.5}_{-0.4}$ & $5.3^{+11.4}_{-2.4}$ &
$1.2^{+0.5}_{-0.3}10^{-4}$ &
0.93/78 \\
\tableline

Src 2 (20\arcsec)/bkg2 & XMM 2000 A&  955/418/470 &
$0.7^{+0.3}_{-0.2}$ & $3.5^{+1.3}_{-0.8}$ &
$1.9^{+0.5}_{-0.3}10^{-4}$ &
1.1/107 \\
Src 2 (20\arcsec)/bkg2 & XMM 2006 B&  --/944/1003 &
$1.0^{+0.4}_{-0.4}$ & $2.2^{+1.2}_{-0.6}$ &
$8.6^{+3.1}_{-2.6}10^{-5}$ &
0.98/105 \\
Src 2 (20\arcsec)/bkg2 & Chandra 2004 & 1829 & $0.9^{+0.3}_{-0.2}$ &
$3.1^{+1.1}_{-0.7}$ & $1.8^{+0.4}_{-0.3}10^{-4}$ &
1.2/94 \\
Src 2 (10\arcsec)/bkg1 & Chandra 2004  & 1270 & $0.7^{+0.2}_{-0.2}$
& $3.7^{+1.3}_{-0.8}$ & $1.5^{+0.3}_{-0.2}10^{-4}$ &
1.4/66 \\

\tableline
Src 3 (20\arcsec)/bkg2 & XMM 2000 B&  --/235/225 &
$0.3^{+2.6}_{-0.3}$ & $7.2^{+72.7}_{-6.0}$ &
$1.9^{+5.4}_{-1.9}10^{-5}$ &
1.2/26 \\
Src 3 (20\arcsec)/bkg2 & XMM 2006 A&  2730/940/1080 &
$0.4^{+0.3}_{-0.2}$ & $36.7^{+42.2}_{-22.8}$ &
$4.0^{+1.0}_{-0.6}10^{-5}$ &
1.2/234 \\
Src 3 (20\arcsec)/bkg2 & Chandra 2004 & 1251 &
$0.4^{+0.4}_{-0.3}$ & $8.1^{+71.8}_{-4.1}$ &
$3.9^{+1.2}_{-0.9}10^{-5}$ &
1.1/66 \\
Src 3 (10\arcsec)/bkg1 & Chandra 2004  & 620 & $0.6^{+0.3}_{-0.2}$ &
$4.3^{+3.7}_{-1.4}$ & $4.8^{+1.2}_{-0.9}10^{-5}$ & 1.3/34
\enddata
\tablenotetext{a}{$N_H$ is in units of 10$^{22}$~cm$^{-2}$.}
\tablenotetext{b}{ Mekal model normalization parameter:
$K=\frac{10^{-14}}{4\pi D^2(1+z)^2} \int n_e n_H dV$,
where D is the angular size distance to the source in cm,
$n_e$ and $n_H$ is the electron and hydrogen density in $cm^{-3}$,
integration is done over source volume.}
\tablecomments{`A' means
PN, MOS1, MOS2 data combined, while `B' means MOS1 and MOS2 data
combined; `bkg1' denotes a background annulus with inner radius
of 10\arcsec and outer radius of 20\arcsec, `bkg2' denotes a
background annulus with inner radius of 20\arcsec and outer
radius of 30\arcsec. The source counts in the third column are given
as PN/MOS1/MOS2 for XMM and ACIS for Chandra. All errors quoted in
the table are at the 90\% confidence level for one interesting
parameter. Src~1* is the annulus region of Src 1(20\arcsec)
with the removed 11\arcsec radius circle containing Srcs 1a and 1b.}
\end{deluxetable*}
%\clearpage

\section{Discussion}

%%%%%%%%%%%%%%%%%%%
Below we discuss the multiwavelength
data obtained for the $\sim$ $1\farcm5$ region that includes Src\,1
and Src\,3  and some implications of the data analysis. We will
refer to this extended source as J0618.
%%%%%%%%%%%%%%%%%%%

The results of the multiwavelength data analysis presented above can
be summarized as follows. The hard extended X-ray source Src~1
consists of `clumps', both extended (Src~1a) and point-like
(Src~1b), which emit hard X-ray continuum. There are signatures of a
thermal component of temperature $\sim$ 0.2~keV in the spectrum of
Src~1a, and of X-ray line emitting clumps side by side with an
extended IR source. The X-ray morphology of the extended source
J0618 is complex. Src~3 is point-like
according to the \chan\ observations. The \xmm\ image in Figure
\ref{chan_xmm} shows evidence for an extended bridge of diffuse
emission connecting Src~1 with Src~3. Longer observations are needed
to establish firmly the possible presence of this faint bridge and of
an X-ray halo around Src~3 (see e.g. Figure \ref{vla2}).

The X-ray spectra of Srcs~1a and 1b and Src~3 are characterized by
power-law components with photon indices $\Gamma \sim$ 1.5. The
spectrum of the Src~1 region shows a feature at about 1.8 keV with a
flux of a few times 10$^{-7} \lfl$ at the 90\% confidence level,
which might be attributed to a Si K-shell line. A feature at 3.7 keV
was found in the X-ray spectrum of Src~3 at the 99\% confidence
level, which might be attributed to an Ar K-shell line, unless the
line is a redshifted Fe K  line of an extragalactic source. A firm
detection of the lines from the localized clumps with \xmm\ is
hampered by the presence of strong hard non-thermal continuum
emission.

An extended source of IR emission was found at the North-Western
edge of Src~1 with a dereddened flux in the 24$\,\mu$m band of {\sl
Spitzer} MIPS of about 3$\times$10$^{-13}$ \efl\ and an upper limit
for the 70$\,\mu$m band of about 10$^{-11}$ \efl. The near-IR flux
of the source is about 2$\times$10$^{-12}$ \efl\ in the 2MASS K$_s$
band.  The source is not seen in the J and H bands of the 2MASS
survey, nor in the POSS-II optical bands. The upper limits are
$3\times$10$^{-13}$ \efl\ for the H band, $7\times$10$^{-13}$ \efl\
for the J band, $3\times$10$^{-11}$ \efl\ for the POSS-II
infrared band (0.7--0.9 $\mu$m), $3\times$10$^{-12}$ \efl\
for the POSS-II red band (0.6--0.7 $\mu$m), and $8\times$10$^{-11}$ \efl\
for the POSS-II blue band (0.4--0.6 $\mu$m).
These upper limits will be used
below to constrain the model we propose for J0618.

Earlier studies of molecular emission from the extended region of
apparent interaction of the IC~443 SNR with the neighboring
molecular cloud have indicated the presence of emission from both
fast and slow shocks (e.g. Burton\etal 1988; Dickman\etal 1992; van
Dishoeck et al.\ 1993, Snell\etal 2005). Molecular clouds are known
to have highly inhomogeneous internal structure (e.g. Blitz 1993). A
molecular cloud consists of numerous dense clumps with a rather
small volume filling factor, embedded in an interclump matter of a
modest density of 5--20 $\cmc$. The presence of a wide range of
dense molecular emission clumps down to about 1\arcsec scale has
been established in the cloud around Src~1 by Richter, Graham, and
Wright (1995).

The apparent coincidence of the extended structured non-thermal
X-ray continuum source J0618 with  local excesses of IR emission can
be understood in the framework of a model of interaction of a
molecular cloud with a fast ballistically moving object. If the
object was ejected by a SNR, it could be either a massive fast
moving ejecta knot or a pulsar wind nebula. In both cases shock
waves will be driven into the cold matter of the cloud. Such a
scenario and the expected properties of the emitted IR and X-ray
radiation will be discussed below. A large power [above
10$^{35}$ erg s$^{-1}$] released by a fragment of velocity about 300
km s$^{-1}$ being decelerated by a dense cloud is emitted as UV photons. The
UV emission will create an HII region surrounding the fragment. The
detected {\sl Spitzer} emission and all of the IR/optical upper
limits are consistent with the continuum emission of an HII region
of a temperature about 10$^4$ K and a number density above 100
cm$^{-3}$. However, to explain the detected near-IR flux of the
source in the 2MASS K$_s$ band one should consider an additional
emission component most likely due to line emission of shocked
molecular hydrogen.
The radio, IR, and optical emission of
the HII region surrounding J0618 will be addressed
in Sections \ref{sect_ene} and \ref{sect_ircont}.

\subsection{X-ray emission from isolated fast ejecta fragments}

An important distinctive feature of J0618 is the presence of
extended non-thermal hard X-ray emission. This can be explained in
the framework of a model of interaction of a massive isolated ejecta
fragment moving with a velocity above 300$\kms$ through a molecular
cloud. In the model by Bykov (2002, 2003), X-ray emission from an
ejecta fragment of 0.2 -- 0.3 pc size ($\sim 30\arcsec - 40\arcsec$
at 1.5 kpc, the angular size of such a fragment in IC~443),
interacting with a molecular cloud was estimated. The fragment mass
was assumed to be $\gsim 10^{-2} \Msun$, containing $\sim10^{-4}
\Msun$ of Si. The ``knot'' traveled through an inter-clump medium of
a number density $\sim 100 \cmc$ with a velocity of $\sim 500\kms$
(the corresponding postshock temperature is $\sim 0.3$ keV). The
model predicts a hard continuum emission with a photon index $\Gamma
\lsim 1.5$ and 1.8~keV Si line emission with a flux of $\sim
10^{-6}~\lfl$. The K-shell lines are excited by both non-thermal and
thermal electrons.

In the case of J0618, a slightly more massive fragment highly
structured due to interaction with the dense molecular clump
would be more realistic. A range
 of sub-fragment velocities around the mean
value $\sim$ 200--300 $\kms$ is consistent with the age of IC~443 of
about 30,000 year advocated by Chevalier (1999). Such a fragment
would provide line fluxes of a few times $10^{-7}~\lfl$, consistent
with the estimated fluxes of the putative Si and Ar lines
(Figure~\ref{fig:Ar}) from J0618. The effects of clumping of the
metal-rich ejecta would result in an intermittent spatial structure
of X-ray line emission from the source. It is worth to mention that
the intrinsic absorbing column $\Delta$\nh\ of a fragment could be
substantial ( $> 10^{21} \cms$), especially if the fragment contains
metal-rich ejecta material.

The relatively low velocity and the high absorbing column of an
ejecta fragment in a dense molecular cloud makes its observational
appearance to be very different from that observed in Vela shrapnel
A. The observed emission of Vela shrapnel A is strongly dominated by
an optically thin thermal component (of $T$ about 0.5 keV) of a
shock heated plasma (Miyata \etal 2001). Contrary to the Vela case,
a thermal emission of a thin plasma of temperature $\sim$ 0.1 keV is
not the main component in the observed X-ray spectra of J0618
because of the high absorption. A hard non-thermal emission
dominates the X-ray spectrum of J0618.

\subsection{Energetics of an X-ray emitting SN ejecta fragment}
\label{sect_ene}

The models of an isolated X-ray source involving the interaction of
a supernova blast wave with dense ambient matter suggest that the
keV emission is due to bremsstrahlung of shock-accelerated electrons
(Bykov\etal 2000; Bykov 2002). Synchrotron X-ray emission would
require electrons with TeV energies accelerated by shocks of speed
well above 1,000 km s$^{-1}$, which is hard to expect for a
middle-aged SNR in a dense ambient medium, unless the moving object
is a pulsar wind nebula.

The radiative efficiency of nonthermal bremsstrahlung is known to
be low at keV energies. This means that to produce a hard X-ray
continuum at a rate $ \dot{\epsilon}_{\rm r}$, an electron of
energy $E_{\rm e}$ dissipates energy at a rate
$\dot{\epsilon}_{\rm e}$ via Coulomb losses in a medium of an
average charge $Z$, and $ \dot{\epsilon}_{\rm
r}/\dot{\epsilon}_{\rm e} \approx 6.3 \times 10^{-4}~ Z (E_{\rm
e}/m_{\rm e}c^2)$ --- see e.g. Akhiezer \& Berestetsky (1957).
Therefore, to produce the X-ray emission at the observed level of
about $6\times 10^{31}\, d_{1.5}^2\, \ergs$ by electrons
accelerated to 50--100 keV, the dissipated power must be about
$\dot{\epsilon}_{\rm e} \approx 10^{35}\, d_{1.5}^2\, Z^{-1}
(m_{\rm e}c^2/E_{\rm e}) \ergs$. The total dissipated power
$\dot{\epsilon_{\rm kin}} = 10 \dot{\epsilon_{\rm e}} \eta_{\rm
-1}^{-1}$. Here $\eta_{\rm -1} \equiv \eta/0.1 \lsim 1$ is the
efficiency of electron acceleration by a shock, providing
$\dot{\epsilon}_{\rm kin} \sim
10^{37}\,d_{1.5}^2\,Z^{-1}\,\eta_{\rm -1}^{-1} \ergs$.The electron
energy $E_{\rm e}$ = 50 keV was assumed in the estimation. Since
$Z \lsim 10$ in an oxygen/silicon-rich gas an enhanced metallicity
of an ejecta fragment (dependent on mixing of the ejecta with the
ambient matter) could somewhat compensate the bremsstrahlung
inefficiency providing $\dot{\epsilon}_{\rm kin} \sim 10^{36}\,
d_{1.5}^2\, \ergs$. Note that the power required to produce
radio-emitting relativistic electrons in clump D is also just
above $10^{36}\, d_{1.5}^2\, \ergs$.

The upstream ram pressure power dissipated at the forward shock of
a ballistically moving fragment, $\dot{\epsilon}_{\rm sh} \approx
8\times10^{35}\, n_{\rm a3} v_{\rm 2}^3\, r_{\rm 20}^2 \,
d_{1.5}^2\, \ergs$, is the source of gas heating and particle
acceleration. Here $r_{\rm 20}$ is the shock radius of J0618
(measured in 20\arcsec). The mechanical power $\dot{\epsilon}_{\rm
kin}$ in the shock model must not exceed $\dot{\epsilon}_{\rm
sh}$, resulting in the condition $n_{\rm a3} v_{\rm 2}^3\, r_{\rm
20}^2\,Z \,\eta_{\rm -1} \geq 12$ to be fulfilled.
Therefore, ejecta fragments of velocity $v_{\rm 2} \sim 2$ moving
through a cloud of density $n_{\rm a3} \sim 0.2$
will have $\dot{\epsilon}_{\rm sh} \approx \dot{\epsilon}_{\rm
kin}$, if $Z \gsim 2$. Thus, the observed patchy structure of the
X-ray emission that is apparent in the \chan\ ACIS images of J0618
could be attributed to structured metal clumps of the ejecta of
IC~443.

How can one directly detect or constrain the power dissipated by the
shock in the model of IC~443 ejecta fragment ballistically moving
through a molecular cloud? The gas temperature behind the standard
single-fluid strong MHD shock can be estimated as $T \approx 1.4
\cdot 10^5~v_{\rm 2}^2~(K)$. For a shock with energetic particle
acceleration efficiency $\gsim$10\% that we consider here the
postshock temperature should be reduced by a factor of about 1.2 to
account for the effects of the energy flux carried away with
energetic particles (see e.g. Bykov 2002). The postshock gas cooling
distance estimated by Hartigan, Raymond, and Hartmann (1987) is about
1.8$\times$10$^{13}~v_{\rm 2}^{4.67}~n_{\rm a3}^{-1}$ cm. The
shock of a velocity 300 $\kms$ in the molecular cloud is radiative
with the cooling layer angular size of 0.15\arcsec $n_{\rm
a3}^{-1}$. If the extended structure of Src 1a seen in Figure~2 of
BBP05 is indeed due to the thermal emission of hot postshock gas
then $n_{\rm a3}\sim$ 0.03 would explain both the extension of about
5\arcsec and the luminosity of soft thermal component below
10$^{32}~\ergs$ discussed in BBP05. Most of the shock power,
however, is not in the soft X-ray emission of shock heated gas, but
rather in UV--optical emission of the radiative shock dominated by UV
lines of OVI at 1035 \AA, Ly$_{\alpha}$ and He II at 304 \AA (e.g.
Hartigan, Raymond, and Hartmann 1987).

At the same time the UV photons produced by the radiative bow shock of ejecta
fragment with a luminosity $L_{UV} = 10^{36}~erg~s^{-1} L_{\rm 36}$,
will be absorbed and reprocessed, mostly to IR emission, in an
expanding HII region surrounding the bow shock of J0618. Note here
that the shape of the HII region could be different from that of the
non-thermal X-ray nebulae, but it should overrun the bow shock of
J0618. From the apparent position of J0618 in the molecular cloud
and the estimated fragment velocity one may conclude that it entered
the molecular cloud about 1,000 years ago. Assuming that all of the
UV photons produced by the radiative bow shock are absorbed in a
spherical layer of an homogeneous ambient matter, one can obtain the
estimation of the HII region radius of about 6$\times$10$^{17}
L_{\rm 36}^{1/3}~ n_{\rm a3}^{-2/3}$ cm (see e.g. Spitzer 1978).  In
fact, the geometry of J0618 is more complex than a spherical HII
region because of the motion of the extended emitting fragment and
because of the strong inhomogeneity of the ambient molecular cloud
down to the arcsecond scale as was observed by Richter, Graham, and
Wright (1995). In the interclump matter of density below 100$\cmc$
the scale size of the HII region would exceed the apparent X-ray size of
J0618. On the other hand, the number density
of the clump D located in a close vicinity of J0618 was estimated by
van Dischoeck \etal (1993) to be above 10$^3 \cmc$ and the size of
the HII region is consistent with the size of the 24 $\mu$m emission
excess around clump D [of about (2--3)$\times$10$^{-11}~\efl$]
apparent in {\sl Spitzer} image presented in the right panel
of Figure~\ref{vla2}. Although very simplified, such a model allows us to
estimate the radio, IR, and optical emission of the shock-produced HII
region.

\subsection{Radio--IR--optical continuum and line emission of the HII region}
\label{sect_ircont}
 The emission of the HII region consists both of continuum emission of a thin
thermal plasma of a kinetic temperature about 10,000 K and of a rich
emission line spectrum (see e.g. Spitzer 1978). The specific
appearance and the chemistry of the mostly neutral photodissociation
region in molecular clouds were reviewed by Hollenbach and Tielens
(1999). In the standard case the interstellar HII regions are
powered by  massive luminous stars. In the case of J0618 the source
of the ionizing radiation is an extended fast moving bow shock and
the ionized region is likely unsteady.  That makes an accurate
modelling of the system rather complicated, so we shall present here
some approximate estimations of the expected IR/optical fluxes from
the HII region.
The observed 24~$\mu$m {\sl Spitzer} MIPS, 2MASS K$_s$ and 1.4 GHz
radio emission, as well as
the upper limits provided by the other {\sl Spitzer}, 2MASS
and POSS-II observations of J0618
were modelled as continuum emission of a hot ionized plasma of
a temperature ranging from 8,000 to 20,000 K.
The modelled fluxes were corrected for interstellar extinction for
a wide range of A$_V$ values below 8, since A$_V \sim$ 8 roughly
corresponds to \nh $\sim$ 1.5$\times$10$^{22}$ cm$^{-2}$ -- the
maximal value allowed by the fits to the Chandra data presented in
Fig.~\ref{fig:CL}. Thus all the optical and IR measurements of
J0618 were used to constrain both the kinetic temperature of the HII
region and the parameter $Y = n_{\rm e3}^2 \times r_{\rm 20}^3$,
where $n_{\rm e3}$ is the electron number density in the HII region
measured in units of 10$^3$ cm$^{-3}$, and $r_{\rm 20}$ is defined in Section~\ref{sect_ene}.
It was found that the measured 24~$\mu$m flux of the source can be explained by a thermal
continuum of a 10,000 K temperature HII region
and $Y \sim$ 0.03 with A$_V \gsim$ 6. Since the continuum emissivity
is $\propto n_{\rm e3}^2$ the apparent patchy image of 24~$\mu$m
emission could be due to the clumpy structure of the ejecta
fragment.

The continuum emission from the HII region is consistent with the
upper limits obtained from {\sl Spitzer}, 2MASS and POSS-II
observations of J0618. The radio flux density from such an HII
region is about 30 mJy. The flux density obtained by integrating the
1.4 GHz radio emission over a larger (an arcminute scale size)
region with VLA was about 60 mJy. Therefore the 1.4 GHz radio flux
is not in a conflict with the HII region model for the parameter $Y
\sim$ 0.03. We found also that the flux of the hydrogen
recombination H$_{\alpha}$ line from the HII region is below the
upper limit of $3\times$10$^{-12}$ \efl\ obtained from the POSS-II
red band (0.6--0.7 $\mu$m) observation discussed above.

The estimated parameter $Y \sim$ 0.03 corresponds roughly to $n_{\rm
e} \sim$ 160 $\cmc$. The bow shock radius estimated from X-ray image
is $r_{\rm 20} \sim$ 1 implying that the derived electron density in
the bright part of the HII region in J0618 is dominated by the
compressed postshock flow. Then the preshock ambient density in the
radiative shock can be estimated from $n_{\rm a} < n_{\rm e}/4$,
providing the ambient gas density $\lsim 40 \cmc$, generally
consistent with that expected in the interclump matter of a
molecular cloud. A substantial part of the powerful UV emission from
the radiative shock of  J0618 will irradiate the nearby dense
molecular clump D providing another HII region of a surface
brightness and high luminosity well in excess of that from J0618
that is clearly seen in Fig.~\ref{vla2}. In the frame of the ejecta
fragment model of J0618 we estimated that at least a substantial
amount of 24~$\mu$m flux from clump D [that is about (2--3)$\times$10$^{-11}~\efl$]
can be attributed to the HII region excited
by the UV emission from the nearby source J0618.

That model provides, however, only about one third of the observed
2MASS K$_s$ band flux of J0618, and that can not be simply relaxed
with an appropriate HII region parameters choice. Thus,
an extra contribution
in the
2MASS K$_s$
band at the level of $\sim 1.5\times$10$^{-12}$ \efl\
is required.
That contribution may come, most likely, from the emission lines of the
shocked molecular hydrogen in J0618
and would require the presence of C-type molecular shocks of
velocity $\sim$~30~km~s$^{-1}$ in the close vicinity of J0618.
Moreover, the atomic fine-structure lines could contribute to the
 24~$\mu$m emission detected by {\sl Spitzer} MIPS if supernova
ejecta drive J-type shocks of about 100~$\kms$ (and faster) into
dense molecular clumps. We discussed above only the emission
produced by the forward bow shock of the fragment, however slower
reverse shock will also be present and since the ejecta fragment
body is likely very inhomogeneous the real structure is likely even
more complex. Relevant line emission models will be briefly reviewed
in the next subsection.

\subsection{IR-line emission in radiative shock models}

Atomic fine structure lines of [OI] (63 $\mu$m) and [FeII] (26
$\mu$m) could dominate the emission in the {\sl Spitzer} MIPS
70$\,\mu$m and 24$\,\mu$m bands. The lines are known to trace fast
radiative shocks in molecular cloud material. A comprehensive study
of radiative shocks in interstellar clouds has been done by
Hollenbach and McKee (1989; HM89 hereafter). They have shown that
the intensity ratio of [OI](63 $\mu$m)/[FeII](26 $\mu$m) is about 10
for a shock wave in a cloud of a density $n_{\rm cl}=
$10$^3$~$\cmc$. They estimated the [OI] (63 $\mu$m) IR line
intensity as $I_{63} \approx 1.5\times 10^{-3}$ ergs s$^{-1}$
sr$^{-1}$ for a shock of a velocity $v_{\rm sh}$ =150 $\kms$ and
demonstrated that it scales roughly linearly with $n_{\rm cl} v_{\rm
sh}$. The combination of optical line emission of OI at 6300 \AA,
CII (2326 \AA), and OII (3726 \AA) provides the dominating gas
coolant in transparent systems of $n_{\rm cl} \leq$ 10$^4 \cmc$. The
line intensities were obtained by HM89 under the assumption of
standard solar composition with account for the interstellar gas
depletion to dust grains. The IR-optical line intensities from
shocks driven by (and into) metal-rich ejecta could be produced with
lower pre-shock number densities n$_{\rm cl}$ than those modeled by
HM89, and should be re-scaled respectively. This can reduce the
pre-shock density required to match the observed fluxes.

Applying the HM89 model intensities to an IR source of an angular
area $A_{100}$ (measured in units of 100 arcsec$^2$) as discussed in
section 3.2, one obtains a [FeII] (26 $\mu$m) line flux $F_{26}
\approx 2.4 \times 10^{-13} A_{100} n_3 v_2\, \enf$. Here $n_3$ is
the pre-shock number density in units of 10$^3 \cmc$ and $v_2$ is
the shock velocity in 10$^2~\kms$. To reach the 24$\,\mu$m flux of
about $3\times 10^{-13}\ \efl$ as estimated from the {\sl Spitzer}
MIPS data, one would need $n_3 v_2 A_{100}\sim 1$. As the estimated
emission area is $A_{100} \sim 2$, one needs $n_3 v_2 \sim  0.5$ to
account for the observed IR emission. The [FeII](26~$\mu$m) line
could provide a sizeable part of the 24~$\mu$m {\sl Spitzer} MIPS
flux. The associated model flux of the [OI] (63~$\mu$m) line,
$F_{63} \approx 2.4 \times 10^{-12} \, A_{100}\cdot n_3 v_2\, \enf$
is consistent with the upper limit derived above. It is apparent
from Figure~\ref{chan_xmm} that Src~1 is located near the shocked
molecular clump D presented in a map by van Dishoeck, Jansen, \&
Phillips (1993). Typical densities of molecular clumps, studied by
van Dishoeck et al.\ (1993) are $\gsim 10^4 \cmc$. The condition
$n_3 v_2 \sim  0.5$ can be matched by either a slow shock of
velocity $v_2 \sim  0.3$ in the outskirts of clump D, or by a fast
shock of $v_2 \sim 1$ in the interclump medium.

Near-IR metastable lines of FeII at 1.3 $\mu$m and 1.7 $\mu$m have
a rather flat shock velocity dependence above 100$\kms$. The
intensity of the brighter FeII line at 1.3 $\mu$m scales roughly
linearly with $n_{\rm cl}$, and it is about $0.2 I_{63}$. This
estimate is close to, but still consistent with the upper limit of
the flux in the 2MASS J band  discussed in Section 3.2. A near-IR
NI (1.04 $\mu$m) line flux was predicted by HM89 to be at the
level comparable with that of [OI] (63 $\mu$m), and a SI (1.1
$\mu$m) line flux similar to that of FeII at 1.3 $\mu$m.

From the apparent lack of extended emission in the POSS-II red band
(0.6--0.7 $\mu$m) the flux upper limit for OI (6300 \AA) can be
estimated as $3\times 10^{-12}$ \efl\ (assuming the reasonable
extinction to Src~1 as $A_{\rm V} \approx 6$). The flux limit is
below the value predicted by HM89 for the OI(6300
\AA)/[FeII](26~$\mu$m) ratio for a transparent system with shock
velocity $v_2 \gsim $ 1. However, the line ratio is increasing with
shock velocity for $v_2 < 1$, and, for example, a fast J-type (see
Draine 1980 for a definition of J- and C-type shocks) radiative
shock of $v_2 \sim  0.6$ in a medium of $n_3 \sim 1$, could explain
both the observed [FeII] 26~$\mu$m line flux and the derived upper
limit for the OI (6300 \AA) line flux. Note that optical line fluxes
from a radiative shock propagating into oxygen-rich ejecta material
could be reduced by heat-conduction effects (e.g. Borkowski \& Shull
1990).

An important diagnostic IR-line ratio sensitive to the shock
velocity is [NeII](12.8 $\mu$m)/[FeII](26 $\mu$m). The ESO VLT
Spectrometer and Imager for the Mid-Infrared (VISIR) is an optimal
instrument for observations in the two mid-infrared atmospheric
windows: the 8--13 $\mu$m N band with a 19$\farcs$2 $\times$
19$\farcs$2 FOV and the 16.5--24.5 $\mu$m Q band with a 32$\farcs$3
$\times$ 32$\farcs$3 FOV (Lagage \etal 2006). The spectrometer could
detect [NeII](12.8 $\mu$m) predicted by the radiative shock model of
HM89 and the continuum emission from Src~1 HII region, thus
discriminating between the two models.

The model of a single fast radiative J-type shock developed by
HM89 underpredicts the near-IR flux (above $10^{-12}\,\enf$)
observed in the K$_s$ band from the extended source
2MASS~J06180378+2227314, closely associated with the {\sl Spitzer}
MIPS 24 $\mu$m excess (Figure~\ref{spit_map}). Thus, a combination
of fast and slow shocks is needed to explain the IR emission from
J0618. The same conclusion has been made in most of the studies of
molecular emission of the south-western region of IC~443. Burton
\etal (1990) argued that both fast dissociative J-type and slow
C-type shocks must be present in the extended southern cloud of
IC~443 to explain the observed line fluxes [see also Snell\etal
(2005) for a recent discussion]. Analyzing different shock models,
including a time-dependent one, Snell\etal (2005) concluded that
no single-shock model can explain the existing observations, and a
range of shock velocities is required.

\subsection{Shocks in a molecular cloud driven by ejecta fragments}

A wide range of shock velocities in a molecular cloud is required
to explain the IR and X-ray observations of Src~1. Slow molecular
C-type shocks are required to explain the observed 2MASS K$_s$
band emission.

Molecular emission occurs when a fast ejecta fragment collides
with molecular clumps of different densities, $\rho_{\rm cl}$,
producing multiple shocks of velocities $v_s \propto \rho_{\rm
cl}^{-1/2}$. This can explain the shock velocity range of $10 \kms
\leq v_{\rm s} \leq 100 \kms$, required to explain both the fluxes
detected by {\sl Spitzer} MIPS and 2MASS K$_s$ and the upper
limits on the optical lines.

In the considered case the presence of a fast shock of velocity
$\gsim$ 300$\kms$ seems to be required to explain the X-ray data
discussed above. Moreover, it is apparent from
Figure~\ref{xmm_spitzer} that while the 1.4~GHz radio emission is
dim in the south-eastern part of the shocked molecular cloud in
comparison with that at the north-east, there is a localized
excess in the vicinity of molecular clump D adjacent to J0618. If
the radio emission is the synchrotron emission produced by
relativistic electrons in the GeV regime in a likely enhanced
magnetic field of the dense clump D, then the model requires the
presence of a fast shock of velocity well above 100 $\kms$ to
accelerate the relativistic particles in the vicinity of Src~1.

The shock ram pressure estimations for multiple shock models were
discussed by many authors (e.g. Burton\etal 1990; Richter, Graham,
and Wright 1995; Cesarsky\etal 1999; Snell\etal 2005). A fast shock
of velocity $v_{\rm s} \sim $50--100$\kms$ incident on a clump of
a moderate number density $10^3$--$10^4 \cmc$ would have a high
ram pressure $n v^2 \sim 10^7$--$10^8 \cmc\ [\kms]^2$. It seems to
be uneasy to attribute the estimated range of ram pressures to a
uniform SN blast wave colliding with a cloud. A SN blast-wave ram
pressure, as estimated from the properties of shocked X-ray
emitting gas inside the remnant, is smaller. On the other hand,
strong ram-pressure inhomogeneities can be produced if a molecular
cloud is hit by SN ejecta driving the blast wave. The ejecta are
likely to be rather fragmented. Each can be considered as an
ensemble of ejecta fragments with a range of velocities and
densities produced by early ejecta instabilities. The fragments
could provide a wide range of ram pressures in the molecular cloud
(up to 10$^8 \cmc\ [\kms]^2$) driving shocks of different
velocities, which are required to explain the observational data.
Isolated fast moving dense ejecta fragments could penetrate deeper
into the dense parts of the cloud driving the high-pressure shocks
on relatively small scales.

\subsection{Time variability and ejecta fragments statistics}

The lifetime of an isolated ejecta fragment in a dense medium is
an important factor with regard to time variability of the X-ray
emission. A fast moving knot is decelerating due to the
interaction with the ambient gas. The drag deceleration time of a
fragment of velocity $v$, mass ${\cal M}$ and radius ${\cal R}$
can be estimated as $\tau_{\rm d} \approx 100 {\cal M}_{\rm -2}
(n_{\rm a3}\, v_{\rm 2}\,{\cal R}_{\rm -1}^{2}\,)^{-1}$ years.
Here ${\cal R}_{\rm -1} = {\cal R}/(0.1 {\rm pc})$ and ${\cal
M}_{\rm -2} = {\cal M}/10^{-2}\,\Msun$. The number density $n_{\rm
a3}$ of the ambient matter is in units of 10$^3$$\cmc$, and the
fragment velocity $v_{\rm 2}$ in units of 10$^2\kms$.
Hydrodynamical crushing of a fast knot occurs on a timescale
$\tau_{\rm c} \sim \chi^{1/2} {\cal R}\,v^{-1}$ (e.g. Chevalier
1975; Sutherland \& Dopita 1995; Wang \& Chevalier 2001, and
references therein). The density contrast $\chi = \rho_{\rm
k}/\rho_{\rm a}$ ($\rho_{\rm k}$ is the knot density) is of the
order of 1 for a large enough fragment, and $\tau_{\rm d} <
\tau_{\rm c}$ for large enough clumps. To reach its apparent
position in the molecular cloud of IC~443, an isolated ejecta
fragment must be massive enough, ${\cal M}_{\rm -2}\gsim 1$, to
overcome strong drag deceleration in the dense matter, as
discussed below.

Particle acceleration occurs on a few-years timescale if the fast
particle diffusion coefficient does not exceed 10$^{22} \diff$
(see Bykov 2002 for a discussion). Thus, given the fragment
lifetime estimated above, some time variability of the X-ray
emission, both in hard continuum and lines, can be expected on a
few-years timescale (and longer) for a fragment of velocity
$v_{\rm 2} \gsim 3$ in a dense ambient matter with $n_{\rm a3}
\gsim 3$, typical for molecular clumps. Variable X-ray emission
has been recently observed with {\sl Chandra} in a few molecular
clouds in the Galactic Center region by Muno \etal (2007). Using
the above estimates, one can argue that variable X-ray emission
could be produced due to an interaction of metal-rich supernova
ejecta with dense clumps of molecular clouds.

According to the $\log N$ -- $\log S$ distribution of X-ray
emitting knots simulated by the method of Bykov (2003), the
probability of another similarly bright ejecta fragment getting
into the field is rather low because of the short lifetime of the
fragment in a dense molecular clump. However, some smaller and
less massive fragments propagating in the interclump medium could
be seen in the cloud as weak point-like X-ray sources of
luminosities $L_x \lsim 10^{31} \ergs$. In order to firmly
identify the sources as such fragments, a deeper observation is
required.

The hydrodynamic simulations available (Klein, McKee, \& Colella
1994; Wang \& Chevalier 2001) predict a complex irregular structure
of the fragment body due to hydrodynamic instabilities. In this case
the X-ray image would show an irregular patchy structure, instead of
a smooth regular head-tail structure. The patches are due to
emission of dense pieces of the fragmented knot illuminated by the
shock-accelerated energetic particles. The observed morphology of
X-ray emission with bright clumps of a few arcsec size in Src~1,
such as Src~1a and Src~1b, as well as Src~3, can be explained in
that model. It is worth to note that ballistically moving ejecta
fragments could have rather large non-radial velocities (with
respect to the apparent center of SN explosion). The non-radial
velocity components will be substantial if the fragments originate
from ejecta instabilities at relatively late evolution stages.

Thus, one can conclude that the morphology, spectra, and X-ray
luminosity of the J0618 complex are generally consistent with
those expected for a ballistically moving SN ejecta fragment
interacting with a molecular cloud, though it is worth to
discuss also some alternatives.

\subsection{Alternative interpretations of J0618}
\label{altern}

A possible interpretation of the extended hard X-ray source J0618
associating it with an interaction of the ejecta of SNR IC~443 with
the nearby molecular cloud has been discussed above in some detail.
Another SNR-related possible interpretation
is a low luminosity pulsar wind nebula originating either from
IC~443 or from another SNR G189.6+3.3, advanced by BBP05.

Extragalactic sources could also contribute to the observed
appearance of J0618. Spectra the of point-like sources Src 2 and Src
3 could be interpreted as that of AGNs.  Within such an
interpretation the feature at about 3.7 keV in the spectrum of Src 3
could correspond to a redshifted (with 0.6 $< z <$ 1) Fe emission
line. Moreover, one could speculate on the association of the
extended Src~1 with a high-$z$ cluster of galaxies.
The angular size of Src~1 ($<$ 1\arcmin at photon energies  $>$ 2 keV)
would suggest in that case a redshift $z >$ 0.5  (see e.g. Rosati 2004 for a
review of high-$z$ clusters). The thermal fit of the diffuse Src~1*
presented in Table~\ref{tpl_spec} with the temperature about 5 keV is consistent
with the cluster interpretation.

Optical and IR identifications of galaxies (of a few arcsecond scale
size at $z \sim$ 1) in the hypothetical cluster are the most obvious
test to check the interpretation and measure the redshift of the
putative cluster. With the optical data available and given the
substantial absorption (3 $ <A_{\rm V} <$ 6), it is possible to
identify the objects not fainter than m$_{\rm V}$ = 17--19, while
the brightest galaxies in most of the known X-ray clusters at $z
\lsim$ 1 are of m$_{\rm V} \gsim$ 19--20. Thus a dedicated
optical-IR study is needed to confirm or reject the high-$z$ galaxy cluster
interpretation. The observed extension of the 2MASS K$_s$ source
correlated with observed diffuse X-ray emission is not easy to
understand in the cluster of galaxies interpretation; it rather
supports a Galactic SNR-related origin of J0618.

\section{Conclusions}
\label{concl}

The multi-wavelength observations presented here indicate a
possible physical connection of the X-ray source J0618 with the
neighboring IR {\sl Spitzer} and 2MASS sources. That connection,
if real, can be understood in a scenario where J0618 originates in
an interaction of the IC~443 SNR with the adjacent molecular
cloud. The correlation would require the presence of both fast and
slow shocks in the clumpy molecular-cloud material. The X-ray line
features apparent in the spectra of the clumps favor a scenario in
which the shocks are produced by a fast ballistically moving SN
ejecta fragment penetrating into a structured molecular cloud. The
model provides a physical picture coherent with the current
observational data, although alternative scenario cannot be
rejected yet.

Alternatively, Src 1 can be interpreted as a massive X-ray cluster
of galaxies at a redshift $z >$ 0.5. High-resolution arcsecond-scale
observations and fine spectroscopy are required to distinguish between
these very different scenarios.

If the SNR-ejecta interpretation is confirmed by further
observations, the source J0618 in the IC~443 could be a prototype of
a rather numerous population of hard X-ray -- IR sources created by
SN explosions in the dense environments of star-forming regions.
Such sources would be particularly abundant in the Galactic Centre
region.

\acknowledgements We thank the anonymous referee for careful reading
of our paper and useful comments. This investigation is based on
observations obtained with {\sl Chandra}, which is operated by the
Smithsonian Astrophysical Observatory on behalf of NASA, and {\sl
XMM-Newton}, an ESA science mission with instruments and
contributions directly funded by ESA Member States and NASA. This
work is based in part on observations made with the Spitzer Space
Telescope, which is operated by the Jet Propulsion Laboratory,
California Institute of Technology under a contract with NASA. This
research has made use of data obtained from the High Energy
Astrophysics Science Archive Research Center (HEASARC), provided by
NASA's Goddard Space Flight Center. This research has made use of
SAOImage DS9, developed by Smithsonian Astrophysical Observatory.
Some of the calculations were performed at the Supercomputing Centre
(SCC) of the A.F.Ioffe Institute, St.Petersburg. A.M.B. and F.B.
thank ISSI (Bern) for a stimulating team meeting on SNRs.

Support for this work was partly provided by NASA through grants
NNX06AE53G and NAG5-10865 and by RBRF grant 06-02-16844, by the
Russian Leading Scientific Schools grant NSh-9879.2006.2, by the
"Extended Objects in the Universe" program of OFN RAS, and by the
"Origin and Evolution of Stars and Galaxies" program of RAS
Presidium. F.B. acknowledges financial contribution from contract
ASI-INAF I/023/05/0. A.M.K. thanks Yu.A.~Shibanov for a valuable
discussion on reduction of optical and IR images.


\begin{references}

\reference{} Akhiezer, A.I., \& Berestetsky, V.B. 1957, Quantum
Electrodynamics, transl. by Consultants Bureau Inc. (Oak Ridge:
Tech. Information Service Ext.)

\reference{} Albert, J., et al., 2007, ApJ, 664, L87

\reference{} Arce, H.G., Goodman, A.A., 1999, ApJ, 512, L135

\reference{} Asaoka, I., \& Aschenbach, B. 1994, A\&A, 284, 573

\reference{} Aschenbach, B., Egger, R. \& Tr\"{u}mper, J. 1995,
Nature, 373, 585

\reference{} Blair, W.P., et al. 2000, ApJ, 537, 667

\reference{} Blitz, L.  1993, in  Protostars and Planets III, ed.
E.H. Levy \& J.I. Lunine, (Tucson: Univ. of Arizona), 125

\reference{} Bocchino, F., \& Bykov, A.M. 2000, A\&A, 362, L29

\reference{} Bocchino, F., \& Bykov, A.M. 2001, A\&A, 376, 248

\reference{} Bocchino, F., \& Bykov, A.M. 2003, A\&A, 400, 203
(BB03)

\reference{} Borkowski, K.J. \& Shull, J.M. 1990, ApJ, 348, 169

\reference{} Braun, R., Strom, R.J. 1986, A\&A, 164, 193

\reference{} Bykov, A.M. 2002, A\&A,  390, 327

\reference{} Bykov, A.M.  2003, A\&A, 410, L5

\reference{} Bykov, A.M., Chevalier, R.A., Ellison, D.C. \&  Uvarov,
Yu.A., 2000, ApJ, 538, 203

\reference{} Bykov, A.M., Bocchino, F., \& Pavlov, G.G.  2005, ApJ,
624, L41 (BBP05)

\reference{} Burton, M.G. 1987, Quarterly Journal of the RAS, 28,
269

\reference{} Burton, M. G., Geballe, T. R., Brand, P. W. J. L., \&
Webster, A. S. 1988, MNRAS, 231, 617

\reference{} Burton, M.G., Hollenbach, D.J., Haas, M.R., \&
Erickson, E.F. 1990, ApJ,  355, 197

\reference{} Cesarsky, D., Cox, P., Pineau des Forets, G., van
Dishoeck, E.F., Boulanger, F., \& Wright, C.M. 1999, A\&A, 348, 945

\reference{} Chevalier, R.A. 1975,  ApJ,  200, 698

\reference{} Chevalier, R.A. 1999,  ApJ,  511, 798

\reference{} Claussen, M.J., Frail, D.A., Goss, W.M., \& Gaume, R.A.
1997, ApJ, 489, 143

\reference{} Cornett, R. H., Chin, G., \& Knapp, G. R. 1977,  A\&A,
54, 889

\reference{} DeNoyer, L.K. 1979, ApJ,  232, L165

\reference{} Dickman, R. L., Snell, R. L., Ziurys, L. M., \& Huang,
Y.-L. 1992, ApJ, 400, 203

\reference{} Draine, B.T. 1980, ApJ, 241, 1021

\reference{} Esposito, J. A., Hunter, S. D., Kanbach, G., \&
Sreekumar, P. 1996, ApJ, 461, 820

\reference{} Fesen, R.A., \& Kirshner, R.P. 1980,  ApJ, 242, 1023

\reference{} Johnson, R.P. 2006, AIP Conf. Proc., 842, 1010

\reference{} Gaensler, B.M., Chatterjee, S., Slane, P.O., van der
Swaluw, E., Camilo, F., \& Hughes, J.P. 2006, ApJ, 648, 1037

\reference{} Green, D.A. 1986,  MNRAS, 221, 473

\reference{} Hartigan, P., Raymond, J., \& Hartmann, L.  1987, ApJ,
316, 323

\reference{} Hewitt, J.W., Yusef-Zadeh, F., Wardle, M., Roberts,
D.A.,\& Kassim, N.E., 2006, ApJ, 652, 1288

\reference{} Hollenbach, D., \& McKee, C.F. 1989, ApJ,  342, 306

\reference{} Hollenbach, D.J., \& Tielens, A.G.G.M. 1999, Rev.
Mod. Phys., 71, 173


\reference{} Kaastra, J.S. 1992, An X-Ray Spectral Code for
Optically Thin Plasmas, (Internal SRON-Leiden Report, updated
version 2.0)

\reference{} Kawasaki, M.T., Ozaki, M., Nagase, F., Masai, K.,
Ishida, M., \& Petre, R. 2002, ApJ, 572, 897

\reference{} Keohane, J.W., Petre, R., Gotthelf, E.V., Ozaki, M., \&
Koyama, K. 1997, ApJ, 484, 350

\reference{} Klein, R.I., McKee, C.F.,  \& Colella, P. 1994, ApJ,
420, 213

%\reference{} Kraft, R.P., Forman, W.R., Hardcastle, M.J., Jones, C.,
%\& Nulsen, P.E.J. 2007, ApJ, 664, L83

\reference{} Lagage, P.-O., Pantin, E., Durand, G., Smette, A.,
Doucet, C., Belorgey, J., \& Pel, J.-W. 2006, Proc. SPIE, 6269,
626913

\reference{} Leahy, D.A. 2004, AJ, 127, 2277

\reference{} Makovoz, D., Roby, T., Khan, I. \& Booth, H., 2006,
Proc. SPIE, 6274, 10

\reference{} Mewe, R., Gronenschild, E.H.B.M., \& van den Oord,
G.H.J. 1985, A\&AS, 62, 197

\reference{} Miyata, E., Tsunemi, H., Aschenbach, B. \& Mori, K.
2001, ApJ, 559, L45

\reference{} Morrison, R., McCammon, D., 1983, ApJ, 270, 119

\reference{} Mufson, S.L., McCollough, M.L., Dickel, J.R., Petre,
R., White,R., \& Chevalier, R. 1986, AJ, 92, 1349

\reference{} Muno, M., Baganoff, F.K., Brandt, W.N., Park, S., \&
Morris, M.R. 2007, ApJ 656, L69

\reference{} Olbert, C.M., Clearfield, C.R., Williams, N.E.,
Keohane, J.W., \& Frail, D.A. 2001, ApJ, 554, L205

\reference{} Petre, R., Szymkowiak, A.E., Seward, F.D., \&
Willingale, R. 1988, ApJ, 335, 215

\reference{} Preite-Martinez A., Feroci, M., Strom, R.G., \&  Mineo,
T. 2000, AIP Conf. Proc., 510, 73

\reference{} Reich, W., Reich, P. \& F\"urst, E. 1990, A\&AS, 83,
539

\reference{} Reid, I.N. et al. 1991, PASP, 103, 661

\reference{} Rho, J., Jarrett, T.H., Cutri, R.M., \& Reach, W.T.
2001,  ApJ, 547, 885.

\reference{} Richter, M.J., Graham, J.R., \& Wright, G.S. 1995, ApJ,
454, 277

\reference{} Rieke, G.H., et al. 2004, ApJS, 154, 25

\reference{} Rosati, P.  2004, in Clusters of Galaxies, ed. J.S.
Mulchaey\& A.Dressler, A.Oemler, (Cambridge University Press), 72

\reference{} Schlegel, D.J., Finkbeiner, D.P., Davis, M., 1998, ApJ,
500, 525

\reference{} Skrutskie, M.F. et al., 2006, AJ, 131, 1163

\reference{} Snell, R.L., Hollenbach, D., Howe, J.E., Neufield,
D.A., Kaufman, M.J., Melnick, G.J., Bergin, E.A. \& Wang, Z. 2005,
ApJ, 620, 758

\reference{} Spitzer, L. 1978,  Physical processes in the
interstellar medium, New York Wiley-Interscience
%, 333 p.

\reference{} Stetson, P.B. 1987, PASP, 99, 191

\reference{} Sturner, S., Keohane, J., \& Reimer, O. 2004, Adv. Sp.
Res., 33, 429

\reference{} Sutherland, R.S. \& Dopita, M.A. 1995, ApJ, 439, 381

\reference{} Tauber, J. A., Snell, R. L., Dickman, R. L., \& Ziurys,
L. M. 1994, ApJ, 421, 570

\reference{} Troja, E., Bocchino, F., \& Reale, F. 2006, ApJ, 649,
258

\reference{} Turner, B. E., Chan, K.-W., Green, S., \& Lubowich, D.
A. 1992, ApJ, 399, 114

\reference{} van Dishoeck, E.F., Jansen, D.J., \& Phillips, T.G.
1993, A\&A 279, 541

\reference{} Wang, C.-Y. \& Chevalier, R.A. 2001, ApJ, 549, 1119

\reference{} Wang, Z.R., Asaoka, I., Hayakawa, S., \& Koyama, K.
1992, PASJ,  44, 303

\reference{} Weisskopf, M.C., Karovska, M., Pavlov, G. G., Zavlin,
V. E., Clarke, T. 2007, Ap\&SS, 308, 151

\reference{} Werner, M.W., et al. 2004, ApJS, 154, 1

\reference{} Winkler, P.F. \& Long, K.S. 2006, AJ, 132, 360
\end{references}
\end{document}